\documentclass[12pt]{article}
\usepackage[dvips]{graphicx}
\usepackage{rotating}

\usepackage{epsfig,amsmath,amssymb,verbatim,mathrsfs}
\usepackage[square, comma, sort&compress]{natbib}

\numberwithin{equation}{section}

\def\beq{\begin{eqnarray}}
\def\eeq{\end{eqnarray}}
\def\bea{\begin{eqnarray}}
\def\eea{\end{eqnarray}}

\def\tev{\, {\rm TeV}}
\def\gev{\, {\rm GeV}}
\def\mev{\, {\rm MeV}}
\def\kev{\, {\rm keV}}

\newcommand{\gsim}{\lower.7ex\hbox{$\;\stackrel{\textstyle>}{\sim}\;$}}
\newcommand{\lsim}{\lower.7ex\hbox{$\;\stackrel{\textstyle<}{\sim}\;$}}

\def\stilde{\widetilde}

\newcommand{\newc}{\newcommand}
\newc{\Nc}{N_{c}}
\newc{\CG}{C_G}
\newc{\gp}{g'}
\newc{\stopi}{\stilde t_i}
\newc{\sboti}{\stilde b_i}
\newc{\staui}{\stilde \tau_i}
\newc{\stopj}{\stilde t_j}
\newc{\sbotj}{\stilde b_j}
\newc{\stauj}{\stilde \tau_j}
\newc{\stopI}{\stilde t_1}
\newc{\stopII}{\stilde t_2}
\newc{\sbotI}{\stilde b_1}
\newc{\sbotII}{\stilde b_2}
\newc{\stauI}{\stilde \tau_1}
\newc{\stauII}{\stilde \tau_2}
\newc{\sstop}{s_{t}}
\newc{\cstop}{c_{t}}
\newc{\ssbot}{s_{b}}
\newc{\csbot}{c_{b}}
\newc{\sstau}{s_{\tau}}
\newc{\cstau}{c_{\tau}}
\newc{\Sstop}{s_{2t}}
\newc{\Cstop}{c_{2t}}
\newc{\Ssbot}{s_{2b}}
\newc{\Csbot}{c_{2b}}
\newc{\Sstau}{s_{2\tau}}
\newc{\Cstau}{c_{2\tau}}
\newc{\salpha}{s_\alpha}
\newc{\calpha}{c_\alpha}
\newc{\Calpha}{c_{2\alpha}}
\newc{\Salpha}{s_{2\alpha}}
\newc{\sbetapm}{s_{\beta_\pm}}
\newc{\cbetapm}{c_{\beta_\pm}}
\newc{\Sbetapm}{s_{2 \beta_\pm}}
\newc{\Cbetapm}{c_{2 \beta_\pm}}
\newc{\sbetaO}{s_{\beta_0}}
\newc{\cbetaO}{c_{\beta_0}}
\newc{\SbetaO}{s_{2 \beta_0}}
\newc{\CbetaO}{c_{2 \beta_0}}
\newc{\vu}{v_u}
\newc{\vd}{v_d}
\newc{\seL}{\stilde e_L}
\newc{\smuL}{\stilde \mu_L}
\newc{\seR}{\stilde e_R}
\newc{\smuR}{\stilde \mu_R}
\newc{\suL}{\stilde u_L}
\newc{\sdL}{\stilde d_L}
\newc{\suR}{\stilde u_R}
\newc{\sdR}{\stilde d_R}
\newc{\scL}{\stilde c_L}
\newc{\ssL}{\stilde s_L}
\newc{\scR}{\stilde c_R}
\newc{\ssR}{\stilde s_R}
\newc{\snue}{\stilde \nu_e}
\newc{\snumu}{\stilde \nu_\mu}
\newc{\snutau}{\stilde \nu_\tau}
\newc{\Gpm}{G^\pm}
\newc{\Hpm}{H^\pm}
\newc{\FFbS}{\overline{FF}S}
\newc{\FFbV}{\overline{FF}V}
\newc{\FSS}{F_{SS}}
\newc{\FSSS}{F_{SSS}}
\newc{\FFFS}{F_{FFS}}
\newc{\FFFbS}{F_{\overline{FF}S}}
\newc{\FSSV}{F_{SSV}}
\newc{\FVS}{F_{VS}}
\newc{\FVVS}{F_{VVS}}
\newc{\FFFV}{F_{FFV}}
\newc{\FFFbV}{F_{\overline{FF}V}}
\newc{\Fgauge}{F_{\rm gauge}}
\newc{\DRbarprime}{$\overline{\rm DR}'$ }
\newc{\DRbar}{$\overline{\rm DR}$ }
\newc{\MSbar}{$\overline{\rm MS}$ }
\newc{\Yu}{{\bf Y}_u}
\newc{\Yd}{{\bf Y}_d}
\newc{\Ye}{{\bf Y}_e}
\newc{\Au}{{\bf a}_u}
\newc{\Ad}{{\bf a}_d}
\newc{\Ae}{{\bf a}_e}
\newc{\bm}{{\bf m}}
\newc{\zhol}{Z^{\rm hol}}

\newcommand{\nnmb}{\nonumber}

\newcommand{\lrf}[2]{\left(\frac{#1}{#2}\right)}

\newcommand{\sep}{s_{\epsilon}}
\newcommand{\cep}{c_{\epsilon}}

\begin{document}

\setlength{\baselineskip}{0.2in}



\begin{titlepage}
\noindent
\begin{flushright}
\end{flushright}
\vspace{1cm}

\begin{center}
  \begin{Large}
    \begin{bf}
Abelian Hidden Sectors at a GeV\\
     \end{bf}
  \end{Large}
\end{center}
\vspace{0.2cm}

\begin{center}

\begin{large}
David E. Morrissey$^{a}$, David Poland$^{a}$,
and Kathryn M. Zurek$^{b,c}$\\
\end{large}
\vspace{0.3cm}
  \begin{it}
$^a$ Jefferson Physical Laboratory, Harvard University,\\
Cambridge, Massachusetts 02138, USA\\
\vspace{0.5cm}
$^b$ Particle Astrophysics Center, Fermi National Accelerator Laboratory\\
Batavia, Illinois 60510, USA\\
\vspace{0.5cm}
${^c}$ Department of Physics, University of Michigan\\
Ann Arbor, Michigan 48109, USA
\end{it}\\

\end{center}

\center{\today}

\begin{abstract}

We discuss mechanisms for naturally generating GeV-scale hidden
sectors in the context of weak-scale supersymmetry.  Such low mass
scales can arise when hidden sectors are more weakly coupled to
supersymmetry breaking than the visible sector, as happens
when supersymmetry breaking is communicated to the visible sector
by gauge interactions under which the hidden sector is uncharged,
or if the hidden sector is sequestered from gravity-mediated
supersymmetry breaking.  We study these mechanisms in detail in
the context of gauge and gaugino mediation, and present specific
models of Abelian GeV-scale hidden sectors.  In particular, we
discuss kinetic mixing of a $U(1)_x$ gauge force with hypercharge,
singlets or bi-fundamentals which couple to both sectors, and
additional loop effects.  Finally, we investigate the possible
relevance of such sectors for dark matter phenomenology, as 
well as for low- and high-energy collider searches.

\end{abstract}

\vspace{1cm}

\end{titlepage}

\setcounter{page}{2}


\vfill\eject



\newpage

\section{Introduction}

  In the minimal supersymmetric extension of the standard model~(MSSM),
the scale of electroweak symmetry breaking is determined by and is
on the order of the scale of soft supersymmetry
breaking~\cite{Martin:1997ns}. Similarly, the effective amount of
supersymmetry breaking in other sectors of the theory can
naturally induce gauge symmetry breaking at the corresponding mass
scale.  If the breaking of supersymmetry is communicated
predominantly by gravitational interactions, the scale of
supersymmetry breaking is typically very similar for all sectors
of theory, even if they do not couple appreciably to one
another~\cite{Sugra,Nilles:1983ge}.
However, if supersymmetry breaking is communicated by gauge
interactions~\cite{Dine:1981gu,Dine:1994vc} under which certain
sectors of the theory are uncharged,
a hierarchy among the scales of supersymmetry breaking can arise
between the different sectors~\cite{Hooper:2008im,Feng:2008ya,
ArkaniHamed:2008qn,Zurek:2008qg,Chun:2008by,Baumgart:2009tn,Cui:2009xq,
Cheung:2009qd,Katz:2009qq}.

  A simple and concrete example of the class of scenarios that
we consider in the present work consists of the MSSM augmented
by an additional hidden $U(1)_x$ sector.  Effects proportional to
the scale of supersymmetry breaking can enter into the hidden
sector in several ways.  Supergravity interactions
are always expected to be present, though their size depends on the
gravitino mass $m_{3/2}$, and whether or not there is sequestering
of generic $M_{\rm Pl}$-suppressed
operators~\cite{Randall:1998uk,Giudice:1998xp}.

  The $U(1)_x$ sector can also have a renormalizable coupling to
the MSSM through kinetic mixing with hypercharge.
Such a term can induce an effective Fayet-Illiopoulos term in the
hidden sector when there is a $D$ term for
hypercharge~\cite{Baumgart:2009tn,Cui:2009xq,Cheung:2009qd},
which can cause the hidden gauge group to break.  
For natural values of the gauge kinetic mixing, the symmetry
breaking scale is on the order of a GeV.
In addition, if supersymmetry breaking is communicated to the visible
sector through the SM gauge interactions ({\em e.g.}, assuming gauge or
gaugino mediation), SUSY-breaking effects will be transferred in
turn to parameters in the hidden sector at the messenger scale
in the presence of kinetic mixing. Renormalization group effects
from the visible-sector gauginos can also induce soft parameters
in the hidden sector through the kinetic mixing term.
These effects are naturally less than or on the order of the
GeV scale.

  If singlets are present in the hidden sector, they may also 
communicate supersymmetry breaking if they couple directly to the
messenger sector or to the MSSM.  With such singlets, no kinetic mixing
is necessary to communicate SUSY breaking to the hidden sector.
In this case, the SUSY-breaking scale in the hidden sector can again
be around a GeV.

  All these effects combine to suggest that hidden sectors may be
found around the GeV scale.  Such new sectors are consistent with
current experimental bounds provided they are sufficiently
hidden, which in the case of gauge kinetic mixing corresponds
to $\epsilon \lesssim 10^{-2}$~\cite{Pospelov:2008zw}.
The detailed phenomenology of these hidden sectors and their
cosmological viability, however, depends strongly on the
relative size of the gravitino mass compared with the mass
of the lightest hidden-sector particle (LHP).  If the gravitino
is much lighter and the hidden sector respects R-parity,
the lightest R-odd particle will
typically be long lived and give rise to problematic decays after
nucleosynthesis or be overabundant.  If the gravitino is heavier,
then the lightest R-odd particle will be stable and one still
needs to ensure that it has an efficient annihilation channel.  We
will show that this is indeed possible in the context of simple
hidden-sector models.

  New hidden sectors at a GeV may help to explain some of
the recent, but surprising, hints for dark matter~(DM).  The
positron and electron excesses seen by
PAMELA~\cite{Adriani:2008zr}, ATIC~\cite{Chang:2008zz}, and
PPB-BETS~\cite{Torii:2008xu}, as well as the WMAP
haze~\cite{Finkbeiner:2003im,Dobler:2007wv,Hooper:2007kb}, can
arise from dark matter annihilation, but seem to require an
enhanced annihilation rate today relative to the value yielding
the correct thermal relic density.  This feature can arise from a
low-velocity Sommerfeld enhancement of the annihilation in our
galaxy, which for electroweak-scale dark matter requires
an attractive force with a mediator lighter
than a few GeV~\cite{Cirelli:2008pk,ArkaniHamed:2008qn}.  New
GeV-mass states are also suggested by the observation of an annual
modulation signal at DAMA~\cite{Bernabei:2008yi}.  This can
potentially be explained by the elastic scattering of GeV-mass dark matter
states~\cite{Gelmini:2004gm,Petriello:2008jj,Savage:2008er,Savage:2009mk,
Chang:2008xa,Fairbairn:2008gz},
or by heavier inelastic dark matter whose inelastic splittings and
scattering cross-section emerge naturally through its coupling to
new GeV-mass states
\cite{TuckerSmith:2001hy,TuckerSmith:2002af,TuckerSmith:2004jv,
Chang:2008gd,Cui:2009xq}.
A GeV sector may also help to account for the $511\,\kev$ line
observed by the INTEGRAL~\cite{Weidenspointner:2006nua}
experiment.

  Some of the mechanisms to generate light sectors which we discuss
here have already been used to construct natural supersymmetric MeV
$U(1)_x$ dark sectors, motivated by the possibility of MeV-mass dark
matter~\cite{Hooper:2008im,Boehm:2003hm,Fayet:2004bw,
Pospelov:2008zw,Pospelov:2007mp}.
The difference between the GeV sectors which are the focus of this
study and MeV sectors relevant there is the strength of the
coupling of the hidden sector to SUSY breaking.  In GeV-scale
supersymmetric sectors, typical couplings to the MSSM and SUSY
breaking are of size $10^{-3}\sim$ GeV/TeV; by contrast, MeV
sectors must have weaker couplings of size $10^{-6}\sim$ MeV/TeV.
However, some of the mechanisms we discuss in the present work can
be applied to MeV sectors as well.
Our study also has overlap with Refs.~\cite{Cheung:2009qd,Katz:2009qq}
that appeared while the present work was in preparation.
Where there is overlap, we confirm their results.

  The outline of this paper is as follows.  In Section~\ref{kgmsb}
we investigate various ways to mediate supersymmetry breaking to
GeV-scale hidden sectors.  Using these results, we construct
several concrete Abelian hidden-sector models in
Section~\ref{models}.  In Section~\ref{apps} we investigate some
applications of these models to explain recent hints for dark
matter, and we discuss briefly their collider signatures. Finally,
Section~\ref{concl} is reserved for our conclusions.

\section{Supersymmetry Breaking and the Hidden Sector\label{kgmsb}}

  We begin with an overview of the different ways that
supersymmetry breaking can be mediated to a hidden sector
containing a $U(1)_x$ gauge symmetry.  Throughout this discussion,
we assume that the MSSM feels supersymmetry breaking primarily
through standard model $SU(3)_c\times SU(2)_L\times U(1)_Y$ gauge
interactions in the form of gauge mediation~\cite{Dine:1981gu,Dine:1994vc}
or gaugino mediation~\cite{Kaplan:1999ac,Chacko:1999mi}.
In the context of gauge mediation, we assume further that the messengers
of supersymmetry breaking to the MSSM are not charged under the
hidden-sector gauge group, and that there is no significant direct 
coupling of the hidden sector to the source of supersymmetry breaking. 
Within gaugino mediation, we assume that the hidden-sector fields 
(including the Abelian gauge fields)
are sequestered away from supersymmetry breaking. In both the
gauge and gaugino mediation frameworks, supersymmetry breaking can
be communicated to the hidden sector through a combination of
supergravity interactions, kinetic mixing of the hidden $U(1)_x$
gauge group with hypercharge, bi-fundamentals charged under both
the hidden and visible gauge groups, and singlets that couple to
both sectors.  We describe each of these possible
contributions below.

\subsection{Supergravity Effects}

  A strong motivation for gauge or gaugino mediation of
supersymmetry breaking, relative to generic gravity mediation,
is that these gauge-based mediation mechanisms provide
an explanation for the absence of strong
flavor mixing induced by TeV-scale soft masses.  For this to be effective,
the MSSM soft terms induced by gauge or gaugino mediation must
strongly dominate over those from supergravity couplings.
Even so, residual supergravity effects can
still potentially provide an important contribution to the
suppressed soft terms in a hidden $U(1)_x$ sector, and those that
break $U(1)_R$ symmetry in particular.

  The typical size of supergravity effects in
both the visible and hidden sectors can be described in terms of
the gravitino mass $m_{3/2}$~\cite{Sugra,Nilles:1983ge},
\beq
m_{3/2} = \frac{F}{\sqrt{3}M_{\rm Pl}},
\eeq
where $F$ parametrizes the underlying supersymmetry breaking.
Generic $M_{\rm Pl}$-suppressed operators connecting the source
of supersymmetry breaking to other fields
then give contributions to the soft parameters in all sectors of the
theory on the order of the gravitino mass $m_{3/2}$.
This leads to a quite general statement -- \emph{in the absence of any
other physics connecting the visible and hidden sectors, mass scales in
hidden sectors are generically comparable to the gravitino mass.}

  An exception to this statement occurs when the generic
$M_{\rm Pl}$-suppressed operators mediating supersymmetry breaking are
further suppressed through \emph{sequestering}. Sequestering can arise
either by localizing a hidden sector on a brane away from the
source of supersymmetry
breaking~\cite{Randall:1998uk,Giudice:1998xp} or through conformal
running effects~\cite{Luty:2001zv}.\footnote{For a pedagogical
review of conformal sequestering, see~\cite{Schmaltz:2006qs}. In
addition to allowing for anomaly mediation to dominate, conformal
sequestering effects have also recently been proposed as a
solution to the $\mu/B\mu$-problem in gauge and gaugino
mediation~\cite{Roy:2007nz,Murayama:2007ge,Perez:2008ng}.} When
sequestering occurs, supergravity effects will still mediate
supersymmetry breaking to all sectors of the theory through
anomaly mediation~\cite{Randall:1998uk,Giudice:1998xp}, generating
soft masses on the order of \beq \Delta m^{AMSB}_{1/2} \sim
\frac{g^2}{(4\pi)^2}\,m_{3/2}, \eeq where $g$ represents a
coupling of the corresponding field.  Anomaly-mediated soft terms
do not generate too much flavor mixing~\cite{Allanach:2009ne}, but
can induce unacceptable tachyonic slepton scalar
masses~\cite{Randall:1998uk}.

  In comparison to these supergravity-mediated soft terms,
gauge-mediated soft terms are on the order of
\beq m_{soft}^{vis} \sim \frac{g^2}{(4\pi)^2}\frac{F}{M} \sim
\frac{g^2}{(4\pi)^2}\lrf{M_{\rm Pl}}{M}\,m_{3/2},
\eeq
where $M$ is the messenger mass scale. Thus, gravity-mediated
soft masses in both the visible and hidden sectors on the order
of a GeV can arise for messenger masses close to $M\sim 10^{14}\,\gev$.
This is about as large as possible while still being consistent with
constraints on new sources of flavor mixing (assuming no new
flavor symmetries)~\cite{Gabbiani:1996hi,Feng:2008zza}.

  Within gaugino mediation, generic supergravity effects
are suppressed by sequestering the source of supersymmetry
breaking. The leading soft terms generated are the visible-sector
gaugino masses, on the order of \beq m_{soft}^{vis} \sim
g^2\lrf{M_c}{\Lambda}\frac{F}{M_c} \sim
g^2\lrf{M_c}{\Lambda}\frac{M_{\rm Pl}}{M_c}\,m_{3/2}, \eeq where
$M_c$ is the compactification scale (inverse of the
compactification length) and $\Lambda$ is the higher-dimensional
cutoff scale. The ratio $M_c/\Lambda$ is less than unity and can
be as small as a loop factor.

  Even with sequestering, residual supergravity effects in all sectors
of the theory will arise from anomaly mediation.  These will be at
the GeV scale when $m_{3/2} \sim 100\gev$, corresponding to $M_c
\sim g^2(M_c/\Lambda) M_{\rm Pl}$. Thus, in this case the cutoff
scale must be close to the Planck scale. Note that this counting
assumes that there are no explicit supersymmetric masses $\sim\int
d^2\theta \mu' H H^c$ or holomorphic K\"{a}hler potential
operators $\sim\int d^4\theta H H^c$ in the hidden sector -- if
there are, conformal compensator effects give contributions to
hidden-sector parameters proportional to
$m_{3/2}$~\cite{Giudice:1988yz,Randall:1998uk,Giudice:1998xp}. In
this case we would need $m_{3/2}\lesssim\gev$, which requires a
somewhat lower cutoff, $M_c \lesssim 10^{14}\gev$.\footnote{In
order to maintain a perturbative description, this may require a
mild $O(10 - 100)$ hierarchy between the cutoff of the higher-dimensional 
effective field theory, $\Lambda$, and the higher-dimensional 
Planck scale~\cite{Buchmuller:2005rt}.}

\subsection{Hypercharge Kinetic Mixing and D terms}

If there is a new $U(1)_x$ gauge group, one can write down a
renormalizable supersymmetric kinetic mixing term connecting it to
hypercharge \beq\label{mixing} \mathscr{L} \supset \int
d^2\theta\,\left(\frac{\epsilon}{2}B^{\alpha}X_{\alpha}
+\frac{1}{4}B^{\alpha}B_{\alpha}
+\frac{1}{4}X^{\alpha}X_{\alpha}\right) + h.c. \eeq Such a term
will be generated radiatively when there are fields charged under
both $U(1)_x$ and
$U(1)_Y$~\cite{Holdom:1985ag,Babu:1996vt,Dienes:1996zr,
Abel:2008ai,Ibarra:2008kn}, 
\beq
\Delta\epsilon(\mu) \simeq
\frac{g_x(\mu)g_Y(\mu)}{16\pi^2}\sum_ix_iY_i\;\ln\lrf{\Lambda^2}{\mu^2}
\label{epsilonloop} 
\eeq 
where $x_i$ and $Y_i$ denote the charges
of the $i$-th field, $\Lambda$ is the UV cutoff scale, and the log
is cut off below $\mu \simeq m_i$, where $m_i$ is the mass of the
$i$-th field. This leads to values of the kinetic mixing in the
typical range $\epsilon \simeq 10^{-4} - 10^{-2}$. Conversely, the
kinetic mixing parameter $\epsilon$ can be highly suppressed or
absent if there exist no such bi-fundamentals, if the
underlying gauge structure consists of a simple group,
or in the context of certain string-theoretic 
constructions~\cite{Dienes:1996zr,Abel:2008ai}.

  If a gauge kinetic mixing term is generated, it will communicate
visible-sector supersymmetry breaking effects to the hidden
sector. This is evident if we shift the basis of gauge fields to
eliminate the gauge kinetic mixing,
\beq
V_x \to \cep V_x,~~~~V_Y
\to V_Y - \sep\,V_x, \label{vshift}
\eeq
where $V_x$ and $V_Y$ are the $U(1)_x$ and $U(1)_Y$ gauge multiplets and
\beq
\sep \equiv
\frac{\epsilon}{\sqrt{1-\epsilon^2}},~~~~~ \cep \equiv
\frac{1}{\sqrt{1-\epsilon^2}}.
\eeq
Doing so, one finds that the
visible-sector fields acquire effective $U(1)_x$ charges, \beq
x_i^{eff} = -\sep\frac{g_Y}{g_x}Y_i.
\eeq
The visible-sector fields then act as gauge messengers to
the hidden $U(1)_x$ sector~\cite{Hooper:2008im}.
We will discuss this in more detail below.

  However, a usually more important effect on the $U(1)_x$ sector
comes from the hypercharge $D$ term, which induces an effective
Fayet-Illiopoulos term~\cite{Fayet:1974jb} in the hidden
sector~\cite{Dienes:1996zr,Suematsu:2006wh,Baumgart:2009tn,Cui:2009xq}.
A hypercharge $D$ term arises when the visible-sector Higgs fields
acquire VEVs with $\tan\beta \neq 1$ induced by SUSY
breaking\footnote{EWSB at the supersymmetric level necessarily has
$\tan\beta = 1$ to ensure $D$-flatness~\cite{Batra:2008rc}.} \beq
\xi_Y = -\frac{g_Y}{2} c_{2\beta} v^2. \eeq The hypercharge FI
term can also receive contributions proportional to supersymmetry
breaking when $Tr(Y m^2)\neq 0$.\footnote{ Of course, there could
be a supersymmetric contribution to $\xi_Y$, but this leads to the
hierarchy problem of why $\xi_Y << M_{\rm Pl}$.} This vanishes in pure
gauge mediation with messenger parity, but can be generated if
there are violations of messenger parity~\cite{Dimopoulos:1996ig},
such as can arise if the Higgs fields couple directly to the gauge
messengers~\cite{Dvali:1996cu}.

  Including the effect of the hypercharge FI term, the $U(1)_x$
$D$-term potential is given by
\beq
V_D = \frac{g_x^2}{2}
\cep^2\left(\sum_ix_i|\phi_i|^2 - \frac{\epsilon}{g_x}\xi_Y
\right)^2,
\eeq
where $\phi_i$ represents a hidden-sector scalar
field with $U(1)_x$ charge $x_{i}$, while the hypercharge $D$-term
potential retains its usual form.  Thus, we see that the
hypercharge $D$ term induces an effective FI term for $U(1)_x$.

  When the FI term dominates the dynamics in the hidden sector,
it can induce a hidden-sector VEV
\beq
\left<\phi_i\right> \simeq
\left(\frac{\epsilon\,\xi_Y}{g_x x_{i}} \right)^{1/2}
\eeq
for one or several of the scalars.  This is close to the GeV scale for
natural values of $\epsilon \sim 10^{-4}\!-\!10^{-3}$.
Alternatively, the hypercharge FI term can be thought of as a
contribution to the hidden sector scalar masses in the amount \beq \Delta
m_{\phi_i}^2 = -\epsilon \,c_{\epsilon}^2 g_x x_{i} \xi_Y. \eeq
This contribution will be in addition to any supersymmetric or
soft supersymmetry-breaking scalar masses present in the hidden
sector. Let us emphasize, however, that even though the induced FI
term is generated (in part) by supersymmetry breaking, and can
itself trigger spontaneous supersymmetry breaking in the $U(1)_x$
sector, it is itself a supersymmetric coupling.

\subsection{Little Gauge Mediation}

  If supersymmetry breaking is communicated to the visible
sector by gauge mediation involving only the MSSM gauge
interactions, soft terms will also be generated for the fields in
the hidden $U(1)_x$ sector if there is gauge kinetic mixing. The
typical size of such terms is less than or on the order of
$m_{soft}^{hid} \sim \epsilon\,m_{soft}^{vis}$, which is close to
a GeV for $\epsilon \sim 10^{-3}$ and $m_{soft}^{vis} \sim \tev$.
Following Ref.~\cite{Zurek:2008qg}, we call this mechanism
\emph{little gauge mediation}.

  The hidden-sector soft terms that arise from integrating out
the gauge messengers can be computed diagrammatically or
(to leading order in $F/M$) through the method of analytic
continuation into superspace~\cite{Giudice:1997ni,ArkaniHamed:1998kj}.
We present a derivation using the analytic continuation technique
in Appendix~\ref{appa}.
The leading order result for the soft scalar masses at the gauge
messenger scale $M$ is
\beq m_{\phi_i}^2 = \epsilon^2\,x_i^2
\lrf{g_x}{g_Y}^2 m_{E^c}^2,
\eeq
with all quantities evaluated at the messenger scale $M$.
Let us emphasize that this result holds both for minimal
and general~\cite{Meade:2008wd} gauge mediation scenarios.

  The $U(1)_x$ gaugino soft mass is a bit more involved since the
kinetic mixing in Eq.~\eqref{mixing} also generates an
off-diagonal gaugino kinetic term.  In the basis where the kinetic
mixing appears explicitly and the visible-sector fields carry no
$U(1)_x$ charges, the $(\lambda_Y, \lambda_x)$ gaugino mass matrix
at the messenger scale at one-loop is simply given by
\beq
M_{gaugino} =
\left(
\begin{array}{cc}
M_1&0\\
0&0
\end{array}\right),
\eeq
where $M_1$ is the standard gauge-mediated contribution to
the hypercharge gaugino mass.  That there is no explicit
$U(1)_x$ gaugino mass generated can be understood as an example
of the ``gaugino screening'' effect discussed in
Ref.~\cite{ArkaniHamed:1998kj}.
In the field basis where the gaugino kinetic mixing is eliminated
by the transformation of Eq.~\eqref{vshift}, the mass matrix becomes
\beq
M_{gaugino} =
\left(
\begin{array}{cc}
M_1&-\sep\,M_1\\
-\sep\,M_1&\sep^2\,M_1
\end{array}\right),
\eeq
which clearly also contains a zero eigenvalue.

  As in gauge mediation to the visible sector, hidden-sector
$A$- and $B$-terms are generated at the two-loop level in the
presence of corresponding hidden-sector trilinear or bilinear
couplings. These terms will be generated at the messenger scale
with size $\sim \frac{\epsilon^2}{(4\pi)^4}\,m_{soft}^{vis}$, and
will generally be subdominant relative to renormalization group
effects, which we discuss below.  Note that, together with our
result for the $U(1)_x$ gaugino soft mass, we find that all
$U(1)_R$-breaking soft terms generated by little gauge mediation
in the hidden sector are suppressed by a least a factor of
$\epsilon$ relative to the $U(1)_R$-preserving scalar soft masses.
We will see in the following section that this feature can have
important implications for the hidden-sector phenomenology.

\subsection{Little Gaugino Mediation}

  A GeV-scale hidden sector can also arise in a natural way
if supersymmetry breaking is communicated to the visible sector by
the MSSM gauginos~\cite{Kaplan:1999ac,Chacko:1999mi}.  Gaugino
mediation can arise in a sequestered extra-dimensional scenario
where the MSSM chiral multiplets as well as the entire $U(1)_x$
sector (gauge and chiral multiplets) are confined to a brane, with
supersymmetry breaking confined to a separate brane.  Sequestering
can also be induced by approximately conformal strong
dynamics~\cite{Luty:2001zv}, and spectra similar to that of
gaugino mediation can arise from coupling gauge mediation to a
conformal hidden sector~\cite{Roy:2007nz,Murayama:2007ge}.

  Allowing only the MSSM vector multiplets to propagate in the bulk
between the branes, the leading-order soft terms consist of MSSM
gaugino soft masses, with all other soft terms vanishing up to
small corrections.  The gaugino masses arise from brane-localized higher
dimension operators of the form~\cite{Kaplan:1999ac,Chacko:1999mi}
\beq
\mathscr{L} \supset \int
d^2\theta\,\lrf{M_c}{\Lambda}\frac{1}{M_c} X W^{\alpha} W_{\alpha},
\eeq
where $\Lambda$ is the cutoff of the higher-dimensional theory
and $M_c$ is the compactification scale.
Such operators generate visible-sector gaugino masses at the
compactification scale $M_c$ on the order of
\beq
M_a
\sim g_a^2\lrf{M_c}{\Lambda}\frac{F}{M_c}.
\eeq
In this scenario, the $U(1)_x$ gaugino will have an approximately
vanishing soft mass at the scale $M_c$ provided it is sequestered
on the MSSM brane.  The leading soft terms in the four-dimensional
low-energy effective theory in both the MSSM chiral and $U(1)_x$
sectors will then arise from renormalization group running (which
we describe below), and additional supergravity effects.

  An additional possibility in gaugino mediation that is consistent with
flavor constraints is that the MSSM Higgs chiral multiplets are
allowed to propagate in the
bulk~\cite{Kaplan:1999ac,Chacko:1999mi,Evans:2006sj}. This can be
important for the hidden sector in that it allows for the
possibility of a non-vanishing hypercharge $D$ term generated by
having $(m_{H_u}^2-m_{H_d}^2) \neq 0$ at the compactification
scale.  RG running will then provide an additional contribution to
$\xi_Y$, beyond that induced by the Higgs VEVs.

  In fact, without specifying the complete UV and GUT structure, it
may be possible to simply write down the brane-localized FI-term
operator \beq \mathscr{L} \supset \int
d^2\theta\frac{1}{\Lambda^2} X^{\dagger}X D_{\alpha}B^{\alpha},
\eeq which is gauge invariant and yields an FI term proportional
to supersymmetry breaking \beq \xi \sim \lrf{F}{\Lambda}^2 \sim
M_1^2, \eeq which is the same order as the visible-sector
hypercharge gaugino mass.  While it would be interesting to further
explore which UV structures may generate this operator
(necessarily involving GUT-breaking effects), we postpone this to
future work.

\subsection{MSSM Renormalization Group Effects}\label{rgeeffects}

  In addition to the soft terms generated at the messenger
threshold scale $M$ or compactification scale $M_c$, the
visible-sector states will themselves act as messengers to the
hidden sector.  These effects are captured in the renormalization
group~(RG) equations of the hidden-sector soft parameters.  They 
are particularly important in relation to $U(1)_R$-breaking 
in the hidden sector.  Indeed, a remarkable property of little gauge (and
gaugino) mediation is that it shields the hidden sector from
$U(1)_R$-symmetry breaking by at least a factor of $\epsilon$.  If
there is no source of explicit supersymmetric $U(1)_R$ breaking in
the hidden sector, a light pseudo-$R$-axion can emerge if the
$U(1)_R$ is broken spontaneously by hidden-sector VEVs.  In
addition, it is even possible for hidden sectors to be
approximately supersymmetric prior to gauge-symmetry breaking, in
which case a pseudo-Goldstino can emerge if supersymmetry is
spontaneously broken.  In these situations, RG running (or
possibly supergravity effects) can provide the dominant source of
$U(1)_R$ breaking, thereby setting the mass of the light state.

  The leading RG effects of the visible-sector soft terms on
the hidden sector come from the hypercharge gaugino mass $M_1$.
Soft scalar masses in the hidden sector receive
a contribution~\cite{Martin:1993zk}
\beq (4\pi)^2\frac{d}{dt} m_i^2 = \sep^2 \left( - 8 x_i^2 g_x^2
|M_1|^2 \right)+ \ldots, \eeq where this is in addition to the
standard contributions to the $\beta$-function from hidden-sector
interactions.  This term can lead to an $O(1)$ correction to the
soft masses generated at the messenger scale if there is a
moderate amount of running.

  Perhaps more importantly, RG effects proportional to $M_1$
generate $R$-symmetry breaking $A$- and $B$-terms in the presence
of supersymmetric hidden-sector bilinear and trilinear
interactions.  These are generated as~\cite{Martin:1993zk}
\bea (4\pi)^2\frac{d}{dt}
b^{ij} &=& -\sep^2 M^{ij}
\left[4 (x_i^2 + x_j^2) g_x^2 M_1 \right]+ \ldots\label{brun}\\
(4\pi)^2\frac{d}{dt} a^{ijk} &=& -\sep^2 y^{ijk} \left[4 (x_i^2 +
x_j^2 + x_k^2) g_x^2 M_1\right]+ \ldots\label{arun}, \eea where
the hidden-sector superpotential is taken to be
$W_{hidden} = \frac12 M^{ij} \Phi_i
\Phi_j + \frac16 y^{i j k} \Phi_i \Phi_j \Phi_k$ and the soft
parameters defined as $V_{hidden} \supset -\left(\frac12 b^{ij} \phi_i \phi_j
+\frac16 a^{ijk} \phi_i\phi_j\phi_k + h.c.\right)$.
Note that the mass parameters generated in
this way are suppressed by a factor of $\epsilon$ relative to the
soft scalar masses.

\subsection{Additional Mediator Fields}\label{addmedfields}

  Finally, there may exist additional fields in the low-energy
spectrum which mediate between the visible and hidden sectors.
 Among the many possibilities, we will focus on two cases:
bi-fundamentals charged under both the visible and hidden gauge
groups, and singlets that couple directly to fields in both
sectors.  In the sections that follow, we will construct explicit
examples that realize both cases.

Light bi-fundamentals in the spectrum, charged under both the
visible-sector gauge group and $U(1)_x$, act as gauge messengers
to the hidden sector.  Such bi-fundamentals develop soft masses
from gauge or gaugino mediation, and then pass on this
supersymmetry breaking through gauge loops connecting them to the
rest of the $U(1)_x$ sector.  The relevant diagrams are analogous
to those arising in standard gauge-mediated models (with the
bi-fundamentals as messengers), with the important difference that
the supertrace of the bi-fundamental multiplets may be
non-vanishing.  The corresponding generalization of minimal gauge
mediation is worked out in Ref.~\cite{Poppitz:1996xw}.
Hidden-sector scalar soft masses are generated on the order of
\beq \Delta m_{i}^2 \sim
-\frac{2g_x^4x_i^2}{(4\pi)^4}\,Str(x_{bf}^2M_{bf}^2)\,
\ln\lrf{\Lambda^2}{M_{bf}^2}, \eeq where $x_i$ is the $U(1)_x$
charge, $M_{bf}$ denotes the bi-fundamental mass matrix, and
$\Lambda$ is approximately the scale at which a vanishing
supertrace is restored.  Up to the logarithmic (running)
enhancement, this leads to soft masses for the pure $U(1)_x$
states suppressed by a loop factor relative to the visible sector.

  Bi-fundamentals also can contribute to $A$- and $B$-terms as well
as the $U(1)_x$ gaugino mass, with a necessary condition being
that their soft parameters contain $U(1)_R$ breaking.  For
bi-fundamental couplings of the form
\bea
W &\supset& \mu_F\,FF^c \\
V_{soft} &\supset& m_F^2|F|^2+m_{F^c}^2|F^c|^2
-\left[(B\mu)_F\,FF^c+h.c.\right], \nnmb
\eea the contribution to the $U(1)_x$
gaugino mass at the scale $\mu_F<M_{mess}$ in the limit of
$\mu_F^2 \gg m_{F}^2= m_{F^{c}}^2\gtrsim (B\mu)_F$ is typically on
the order of \beq M_x \simeq
\frac{g_x^2x_F^2}{8\pi^2}\frac{(B\mu)_F}{\mu_F}, \eeq with the
general expression for the gaugino mass given in
Ref.~\cite{Poppitz:1996xw}.  An important special case occurs when
$\mu_F$ is a genuinely supersymmetric threshold (i.e., the 
$(B\mu)_F$ term arises entirely from supersymmetry breaking contained
in the wavefunction renormalization of $F$ and $F^c$).  
With this provision, the gaugino screening theorem of
Ref.~\cite{ArkaniHamed:1998kj} implies that the contributions to
$M_x$ cancel at this loop order, leading to a highly suppressed
gaugino mass.

  Bi-fundamental fields in the spectrum will also induce (or add to)
kinetic mixing between $U(1)_x$ and hypercharge, which in turn
will further mediate supersymmetry breaking to the hidden sector.
In fact, this scenario is continuously connected to the situation
considered previously, where all bi-fundamentals were assumed to
have been integrated out above the scale of the gauge messengers
or the compactification scale.  The only difference is that we
have now lowered the bi-fundamental mass scale below the messenger
scale. Even so, let us also mention that the additional
logarithmic enhancement in $\epsilon$ due to lighter
bi-fundamentals typically requires relatively small values of the
$U(1)_x$ gauge coupling $g_x \lesssim 0.1$ (assuming order unity
charges) to avoid generating an unacceptably large low-scale value
of $\epsilon$.

  Supersymmetry breaking can also be mediated to the hidden sector
by gauge-singlet chiral superfields.  As an example, consider the
gauge singlet $S$ with superpotential couplings
\beq W \supset
\zeta\,S\,H_u\,H_d + \lambda\,S\,H\,H^c.
\label{ncoup}
\eeq
Here, $H_u$ and $H_d$ are the MSSM Higgs multiplets and $H$ and $H^c$
are a vector pair of states charged under $U(1)_x$.  The singlet
$S$ may or may not condense to generate the $\mu$ term. The
interactions of Eq.~\eqref{ncoup} will then generate a running
contribution to the soft scalar masses of $H$ and $H^c$,
\bea
(4\pi)^2\frac{d}{dt} m_S^2 &=&
4|\zeta|^2(m_S^2+m_{H_u}^2+m_{H_d}^2+|A_{\zeta}|^2) \\
&&~~~ +2|\lambda|^2(m_S^2+m_{H}^2+m_{H^c}^2+|A_{\lambda}|^2) + \ldots\nnmb\\
(4\pi)^2\frac{d}{dt} m_{H^{(c)}}^2 &=&
2|\lambda|^2(m_S^2+m_{H}^2+m_{H^c}^2+|A_{\lambda}|^2) + \ldots
\eea
At the messenger or compactification scale, the Higgs soft
masses are typically positive. The singlet soft mass, which
vanishes at the input scale in minimal scenarios, is then driven
negative.  This in turn drives the soft masses for $H$ and $H^c$
positive.  As long as $\lambda$ or $\zeta$ are somewhat small,
this can lead to GeV-scale contributions to hidden-sector
parameters.

  Alternatively, $S$ can couple directly
to the supersymmetry-breaking sector and pick up a large soft mass
that then drives the $H$ and $H^{c}$ masses negative.  This can
in turn cause the hidden-sector gauge group to break.  An additional
possibility is that the $H$ and $H^c$ fields receive weak-scale
masses and supersymmetry-breaking parameters through larger
couplings and possible (auxiliary) VEVs of $H_u$, $H_d$, and $S$.
In this case, they play a similar role as the bi-fundamentals
considered above, mediating supersymmetry breaking to the rest of
hidden sector.

\section{Models\label{models}}

  We present here a handful of simple models for the hidden $U(1)_x$
sector that illustrate the supersymmetry-breaking mechanisms
discussed in Section\;\ref{kgmsb}. In many cases, the models we
consider run into problems with constraints from nucleosynthesis
or generate dark matter in excess of the observed density.
As such, our main goal will be to construct simple viable hidden
sectors that avoid these difficulties.  We will see that, because
of these constraints, it is generally favorable for the gravitino mass
$m_{3/2}$ to be heavier than the lightest hidden-sector particle
(LHP).  We begin with a simple little-gauge-mediated model whose
matter content is a pair of hidden Higgs fields with a bare $\mu'$
term.  We will see that this model generically contains a light
fermion with an overly large relic density unless supergravity
effects push its mass above that of the $U(1)_x$ gauge boson. This
regime also gives a natural explanation of the origin of the
$\mu'$ term, but has the disadvantage that the spectrum depends
on unknown UV physics.

  We then turn to a hidden NMSSM model which promotes the $\mu'$
term to a hidden singlet.  When the gravitino is lighter than the
LHP, the scenario is only viable for a very low
supersymmetry-breaking scale, or if additional operators allow the
LHP to decay.  We will see that making the gravitino heavier
allows for additional viable scenarios as long as the LHP retains
a large enough annihilation channel.  An interesting possibility
for doing this is to take $m_{3/2} \sim m_{hid}$ while
sequestering supergravity effects.  In this case the LHP still has
phase space to annihilate to nearly degenerate gauge bosons, with
the mass splitting set by anomaly mediation effects.  The LHP is
then a viable candidate to be a component of the dark matter.

  Lastly, we discuss models where the mediation occurs via
dynamical matter fields, and construct an explicit model for singlet
mediation and mediation by bi-fundamentals.  The singlet scenario
often contains a light pseudo-axion, but can be viable provided the
axion is able to decay through small couplings to the visible sector.
With bi-fundamentals, the low-energy spectrum and phenomenology
is frequently very similar to the case of little gauge mediation.

\subsection{Minimal $\mu'$ Model}

  The minimal viable $U(1)_x$ dark hidden sector consists of
a vector-like pair of chiral multiplets $H$ and $H^c$ with the
superpotential \beq W \supset \mu'\,H H^c. \eeq The most natural values
of $\mu'$ are either $m_{3/2}$ or 0 (or very large!), and we
include in this class of models the limit $\mu'\to 0$.  To permit
symmetry breaking, $\mu'$ must not be much larger than other
contributions to hidden-sector parameters.  For the time being we
leave the origin of $\mu'$ unspecified, although we will comment
on how values of $\mu' \lesssim \gev$ can arise naturally from
supergravity effects when $m_{3/2}\sim \gev$.  In analyzing this
model, we will assume that $U(1)_x$ has kinetic mixing with
hypercharge as discussed above.

  In the absence of supersymmetry breaking, the tree-level
low-energy scalar potential of the theory is given by
\bea
V &=& |\mu'|^2(|H|^2 + |H^c|^2)  \\
&&+\frac{g_x^2}{2}\left(x_H|H|^2-x_H|H^c|^2 - \xi \right)^2.
\nnmb
\label{upot} \eea The parameter $\xi$ appearing in the $D$-term
part of the potential arises from the hypercharge FI term set by
the VEV of the MSSM Higgs fields, and is given by
\beq
\xi \equiv \frac{\epsilon}{g_x}\,\xi_Y =
-\frac{\epsilon}{g_x}\frac{g_Y}{2}c_{2\beta}\,v^2.
\eeq
This term will receive additional contributions if there is a
non-vanishing FI term for either hypercharge or
$U(1)_x$, or if $Tr(Ym^2)\neq 0$.

  Without loss of generality, we assume $x_H > 0$ and $\epsilon > 0$
so that $\xi$ is positive and the potential gives $H$ a negative
mass-squared contribution.  With these sign conventions, the
minimum of the potential lies at
\beq
\left<H\right> \equiv \eta =
\sqrt{\xi/x_{H}-|\mu'|^2/(g_x x_H)^2},~~~~~ \left<H^c\right> =0.
\eeq
Thus, even in the absence of soft supersymmetry-breaking
terms in the hidden sector, the gauge symmetry (as well as
supersymmetry) is spontaneously broken.

  The corresponding spectrum of bosonic states then consists of
a massive gauge boson and a physical Higgs boson $h$ derived from
$H$, both with mass
\beq m_h^2 = m_{Z_x}^2 = 2g_x^2x_H^2\eta^2.
\label{Zxmass}
\eeq
as well as a complex scalar derived from $H^c$ with mass $2|\mu'|^2$.
For $g_xx_H = 0.1$, $\epsilon = 10^{-3}$, and $\xi$ generated only
by the MSSM Higgs VEVs, we find $m_{Z_x} \lesssim 1\,\gev$.
As $\mu' \to 0$, the potential acquires a flat direction at tree-level
corresponding to this state becoming massless and supersymmetry
in the hidden sector being restored.

  Among the fermionic states, two are comprised of
a Dirac mixture of the Higgsinos and the $U(1)_x$
gaugino with mass
\beq
M_{2,3}^f = \sqrt{|\mu'|^2+m_{Z_x}^2}.
\label{fermionmass}
\eeq
The third fermionic state is a massless Weyl fermion.
It is the Goldstino corresponding to the spontaneous breaking
of supersymmetry in this sector.

  The spectrum will be deformed by the inclusion of soft
supersymmetry-breaking operators. The precise effect of the soft
terms depends on their origin, whether from gauge or gaugino
mediation or due to residual supergravity effects.  In general,
however, the phenomenology of the hidden sector can be classified
according to the scale of the hidden-sector masses relative to the
gravitino mass $m_{3/2}$.  Thus, we consider the distinct cases
$m_{3/2}\ll m_{hid}$ and $m_{3/2} \gtrsim m_{hid}$.

\subsubsection{$\mathbf{m_{3/2}\ll m_{hid}}$}

  Such a hierarchy can arise with a
messenger or compactification scale well below $10^{14}\gev$ in
gauge or gaugino mediation.  Little gauge or gaugino mediation
effects then provide the dominant contributions to the soft terms,
which are parametrically smaller by a power of $\sqrt{\epsilon}$
compared to the symmetry breaking induced by $\xi$.  We can thus
treat their effects as small perturbations on the supersymmetric
spectrum described above.

  The tree-level scalar potential becomes
\bea
V &=& (m_{H}^2+|\mu'|^2)|H|^2 + (m_{H^c}^2+|\mu'|^2)|H^c|^2 \nnmb\\
&&- (B\mu'H\,H^c+h.c.)\\
&& +\frac{g_x^2}{2}\left(x_H|H|^2-x_H|H^c|^2 -\xi \right)^2.\nnmb
\label{upotwsoft} \eea For the time being, we will assume that the
$\mu'$ term is genuinely supersymmetric in origin and neglect
supergravity effects. The $B\mu'$ term is then generated within
little gauge or gaugino mediation primarily through RG running as
in Eq.~\eqref{brun}, on the order of $B\mu' \lesssim
g_x^2\epsilon^2\mu'\,M_1$, and is parametrically smaller than the
scalar soft masses by a factor of $\epsilon$, which in turn are
smaller than the symmetry breaking induced by the FI term by a
factor of $\sqrt{\epsilon}$. This has important implications for
the spectrum and phenomenology of the model.

  With the inclusion of soft terms, the VEV of $H$ is shifted to
\beq
\left<H\right> \equiv \eta = \sqrt{\xi/x_H - (m_{{H}}^2 +
|\mu'|^2)/(x_H g_x)^2}.
\eeq
The $B\mu'$ term forces the $H^c$ field to develop a VEV as well.
Since this term is subleading, the
corresponding VEV of $H^c$ is much smaller than that of $H$.
Denoting the ratio of VEVs by $\tan\alpha =
\left<H^c\right>/\left<H\right>$, we find
\beq
\tan\alpha \simeq
\frac{B\mu'}{m_{H}^2 + m_{H^c}^2 + 2|\mu'|^2}. \label{tanalpha}
\eeq
With our assumption of a supersymmetric origin of the $\mu'$
term, this ratio is on the order of
$\tan\alpha \sim \epsilon^2 M_1 / |\mu'|$.

  The bosonic spectrum still contains a single real Higgs scalar
that is degenerate (at tree-level) with the gauge boson with mass
given again by Eq.~(\ref{Zxmass}),
as well as CP-even and CP-odd scalars derived primarily from
$H^c$ with masses
\beq m_{h^c}^2 \simeq m_{a^c}^2 \simeq m_H^2 + m_{{H}^c}^2 + 2 |\mu'|^2.
\eeq
These states are split slightly by the $B\mu'$ term.

  The fermion mass matrix becomes
\begin{equation}
\mathcal{M}^f
\simeq
\left( \begin{matrix}
0 & -\sqrt{2} x_H g_x \eta \tan\alpha & \sqrt{2} x_H g_x \eta \\
-\sqrt{2} x_H g_x \eta \tan\alpha & 0 & \mu' \\
\sqrt{2} x_H g_x \eta & \mu' & 0
\end{matrix}    \right).
\end{equation}
Two of the mass eigenvalues are relatively heavy, and correspond
to a nearly Dirac state with mass given by Eq.~(\ref{fermionmass}).
The third fermion state is much lighter, and coincides with the
massless Goldstino fermion found above.  Explicit supersymmetry and
$R$-symmetry breaking in the form of the subleading $B\mu'$ term
(or gaugino mass $M_x$) is required to lift this state.  Including
the effect of $B\mu'$, the light mass is
\beq
M_1^{f} \simeq \left( \frac{2m_{Z_x}^2}{m_{Z_x}^2 + |\mu'|^2} \right)
\mu' \tan\alpha
\sim \epsilon^2 g_x^2 M_1.
\eeq
For $M_1 \sim 100\,\gev$, $\mu'\sim \gev$, $\epsilon \sim 10^{-3}$,
this is on the order of $0.1\,\mev$.

  The very light pseudo-Goldstino fermion state is evidently the
lightest hidden particle~(LHP) in this sector.  It is metastable
due to $R$-parity (whose conservation requires $H$ and $H^c$
to be even), and is only able to decay or annihilate into
visible sector states through kinetic mixing.
We estimate the corresponding annihilation cross-section through
$s$-channel gauge boson exchange into $e^+e^-$ ($M_1^f > m_e$)
or $\nu\bar{\nu}$ ($M_1^f< m_e$) to be~\cite{Hooper:2008im}
\beq
\left<\sigma v\right> \simeq \left\{
\begin{array}{ccc}
\frac{g_x^2x_H^2e^2c_W^2}{3\pi}\,|U_{fx}|^4\epsilon^2
\frac{(M_1^f)^2}{m_{Z_x}^4}\,v_{f.o.}^2 \sim \frac{g_x^2
x_H^2e^2c_W^2}{12\pi}\,|U_{fx}|^4\epsilon^6
\frac{M_1^2}{\eta^4}\,v_{f.o.}^2,
&M_1^f > m_e,\\
\frac{g_x^2x_H^2{g'}^2}{4\pi}\,|U_{fx}|^4\epsilon^2
\frac{(M_1^f)^6}{m_{Z_x}^4m_{Z}^4}\,v_{f.o.}^2 \sim
\frac{g_x^{10}x_H^2{g'}^2}{16\pi}\,|U_{fx}|^4\epsilon^{14}
\frac{M_1^6}{\eta^4m_{Z}^4}\,v_{f.o.}^2, &M_1^f < m_e ,
\end{array}
\right.
\eeq
where $v_{f.o.}$ is the typical velocity during freeze-out
and $U_{fx}$ is the gaugino fraction of the state.
The additional suppression for $M_1^f < m_e$ comes from the
fact that the annihilation final state consists of neutrinos.
For $m_{Z_x}^2 \ll m_{Z}^2$, the $U(1)_x$ mixes primarily with electromagnetism
and couplings to neutrinos (due to a residual mixing with the $Z^0$)
are suppressed~\cite{Baumgart:2009tn}.   This cross-section is much
too small to yield an acceptable relic density, even with $M_1^f > m_e$.
Indeed, for $M_1^f < m_e$, the light fermion state decouples
while it is still relativistic.

  Under our assumption of a very light gravitino, the light
fermion is unstable to decay into a photon and a gravitino
via its $\epsilon$-mediated mixing with the photino.
The corresponding lifetime is
\bea
\tau &=& \frac{16\pi\left<F\right>^2}{m_{\chi}^5\,|P_{\gamma}|^2}\\
&\simeq& (3\times 10^{23}\,s)\,
\lrf{\sqrt{\left<F\right>}}{100\,\tev}^4\lrf{0.1\,\mev}
{m_{\chi}}^5\,\lrf{\epsilon}{|P_{\gamma}|}^2\lrf{10^{-3}}{\epsilon}^2\nnmb,
\eeq where $F$ is the auxiliary VEV parametrizing the underlying
source of supersymmetry breaking in all sectors, and $P_{\gamma}
\sim \epsilon$ is the projection of the light $U(1)_x$-sector
fermion onto the photino. This lifetime is on the order of the age
of the universe. Since the relic density of the light state is too
large, this scenario is ruled out as it stands.

  So far, we have completely neglected supergravity contributions to the
hidden-sector soft parameters.  However, even though we have
assumed $m_{3/2} \ll m_{hid}$, we could still have $m_{3/2}^2
\gtrsim g_x^2\epsilon^2|\mu'|M_1$, in which case residual
supergravity effects would provide the dominant contribution to
$B\mu'$ (and $M_x$).  The primary effect would be to push up the
mass of the light fermion, which would weigh nearly the same as
the gravitino. If the light fermion is lighter than the gravitino
it is then stable, and will still generically have too large of a
relic density.  In either case, whether the gravitino is lighter than
the lightest fermion in the hidden sector (in which case the fermion
decays to the gravitino after BBN) or if the gravitino is heavier
than the fermion (in which case the fermion has too high a relic
density), this scenario is not viable.

\subsubsection{$\mathbf{m_{3/2}\gtrsim m_{hid}}$}

  Our results lead us to consider larger values of $m_{3/2}$,
such as might emerge with a messenger or compactification scale on
the order of or greater than $M\sim 10^{14}\gev$ in gauge or
gaugino mediation. Indeed, the case of gauge mediation with
$m_{3/2}\sim 1\,\gev$ arises in \emph{sweet-spot} models of
supersymmetry breaking~\cite{Ibe:2007km}, as well as in certain
GUT constructions based on
$F$-theory~\cite{Beasley:2008dc,Beasley:2008kw,Heckman:2008qt,Marsano:2008jq}.
Residual supergravity-induced flavor-mixing effects in this case
lie on the border of what is consistent with current
constraints~\cite{Feng:2008zza}.  The precise effect on the hidden
sector spectrum for a given value of $m_{3/2}$ depends on whether
or not supergravity-mediated effects are sequestered.  We consider
both possibilities here.

  In the case of gauge mediation without sequestering,
residual supergravity-mediated effects will generate contributions
to all soft masses on the order of $m_{3/2}$. Thus, if the hidden
sector is to be light (and to avoid flavor constraints)
we should have $m_{3/2} \lesssim \gev$.  For $m_{3/2} \sim \gev$
supergravity contributions to the hidden-sector soft terms will
be of the same order as those due to kinetic mixing.
A value of $\mu' \sim \sqrt{|B\mu'|} \sim m_{3/2}$ can also
arise through the standard Giudice-Masiero mechanism~\cite{Giudice:1988yz}.
The precise particle spectrum will therefore depend on unknown UV physics.

  When the lightest $R$-parity odd state in the spectrum
is lighter than the gravitino, it will be a stable dark matter candidate.
If the gauge coupling $g_x$ is somewhat small, supergravity
contributions can push all the gaugino and Higgsino states
that could make up the stable state to be heavier than the
gauge boson and the scalar Higgs $h$.
This allows the lightest $R$-odd fermion to annihilate very
efficiently to gauge bosons, with rate~\cite{Griest:1989zh}
\bea \left<\sigma
v\right> &\simeq&
\frac{g_x^4x_h^4}{64\pi}\frac{|U_{fh}|^4}{{M_1^f}^2}\\
&\simeq& (6\times 10^{-24} \mbox{
cm}^3/\mbox{s})\lrf{1\,\gev}{M_1^f}^2
\lrf{g_xx_H}{0.1}^4|U_{fh}|^4,
\nnmb
\eea
where $U_{fh}$ is an order unity mixing factor into Higgsinos.
One can thus easily avoid too large a relic density for the
stable $R$-odd fermion.

  The gauge boson and the hidden Higgs $h$ will themselves decay rapidly
to the visible sector.  Of these two states, the hidden Higgs $h$
will be longer-lived, with decays to photon pairs by mixing with
the visible SM-like Higgs boson $h^0$, as well as decays to
electrons through kinetic mixing. The lifetime for di-photon
decays through Higgs mixing is on the order of \bea
\tau_{h\to\gamma\gamma} & \simeq &
\frac{256\pi^3}{\alpha^2}\frac{1}{\epsilon^2}
\frac{m_{h_0}^4}{m_h^5}\\
&\simeq&
(2\times 10^{-2}\,s)\,\lrf{10^{-3}}{\epsilon}^2
\lrf{m_{h^0}}{115\,\gev}^4\lrf{1\,\gev}{m_h}^5.
\nnmb
\eea
This is typically longer than the lifetime for hidden Higgs decays to
four electrons via kinetic mixing, though it goes through
a higher power of $\epsilon$:
\bea
\label{hdecay4e}
\tau_{h\to 4e} &\simeq& \frac{256\pi^3}{g_x^2x_H^2e^4c_W^4}
\frac{1}{\epsilon^4}\frac{1}{m_h}\\
&\simeq& (1\times 10^{-4}\,s)\,\lrf{0.1}{g_xx_H}^2
\lrf{10^{-3}}{\epsilon}^4\lrf{1\,\gev}{m_h}.\nnmb
\eea
There is also a related loop-mediated decay to two electrons
that we estimate to be of similar size.  Thus, this scenario leads
to a viable, but UV-dependent, phenomenology.

  When supergravity effects are sequestered, such as is required
for (high-scale) gaugino mediation to be the dominant source of
the MSSM soft masses, it is possible for anomaly-mediated
contributions to set the scale of the hidden sector, with $m_{hid}
\sim m_{3/2}/(4\pi)^2$~\cite{Katz:2009qq}.  However, in order to
avoid generating too large of a hidden-sector $B\mu'$ term, we
must have $\mu' = 0$ in the superpotential and forbid the operator
$HH^c$ in the K\"ahler potential.  In this case, the scenario with
vanishing superpotential is highly predictive assuming there are
no additional non-decoupling effects.  However, it has the problem
that the hidden-sector potential is unstable.  This is because the
soft masses induced by anomaly mediation are given by \beq m^2_H =
m^2_{H^c} = -\frac{4 g_x^4x_H^4}{(4\pi)^4}\;m_{3/2}^2 < 0 \eeq and
so the D-flat direction $\left<H\right>=\left<H^c\right>$ is not
stabilized.  Including an FI term in the potential does not help
stabilize this direction because the sum of the squared soft
masses does not change (and is still negative).  Thus, this
scenario requires additional contributions to the potential in
order to be viable.

\subsection{Hidden Sector NMSSM}\label{hidsecnmssm}

  The minimal model described above contains
a dimensionful coupling $\mu'$ whose origin can be problematic
(unless $m_{3/2}\sim m_{hid})$, and often gives rise to a light
fermion state or an unstable potential.  Both of these possible
difficulties can be overcome by adding a singlet $S$ to the theory
and taking the superpotential to be \beq W \supset
\lambda\,S\,H\,H^c. \eeq We can enforce this form by imposing a
discrete or continuous global symmetry in addition to the gauged
$U(1)_x$. Such a symmetry can also prevent a direct coupling of
the singlet $S$ to the visible-sector Higgs fields, the gauge
messengers, or the supersymmetry-breaking sector.

  Neglecting soft supersymmetry-breaking operators,
the tree-level scalar potential in this theory is
\bea
V &=&
|\lambda|^2\,|H|^2\left(|H^c|^2+|S|^2\right)+|\lambda|^2|H^c|^2|S|^2 \\
&&+ \frac{g_x^2}{2}\left(x_H|H|^2-x_H|H^c|^2-\xi\right)^2,\nnmb
\eea
where again we take $\xi > 0$ and $x_H > 0$.  The
supersymmetric global minimum of this potential has
$\left<H\right> \equiv \eta \simeq \sqrt{\xi/x_H}$, along with
$\left<S\right> = 0 = \left<H^c\right>$. At this minimum, the
theory has an exact global $U(1)$ symmetry under which $S$ and
$H^c$ have opposite charges. As a result, the theory breaks into
two sectors, with the lightest state in the sector derived from
$S$ and $H^c$ absolutely stable up to explicit breaking by
higher-dimensional operators.\footnote{Including an $S^3$ term in
the superpotential breaks the global $U(1)$ down to a
$\mathbb{Z}_3$ subgroup, but still gives rise to a stable state in
this sector.}

  The spectrum of the theory within this supersymmetric minimum
consists of a massive vector multiplet and a pair of chiral multiplets.
This can be seen from the fermion mass matrix, which in the basis
$\left(\tilde{\lambda},\,\tilde{H},\,\tilde{H^c},\,\tilde{S}\right)$
is given by
\beq
\mathcal{M}^f = \left(
\begin{array}{cccc}
0&\sqrt{2}x_H g_x\eta&0&0\\
\sqrt{2}x_H g_x\eta&0&0&0\\
0&0&0&\lambda\,\eta\\
0&0&\lambda\,\eta&0
\end{array}
\right). \label{fermionmassmatrix}
\eeq
The condensing $H$ chiral
multiplet gets eaten by the gauge multiplet yielding a massive
vector multiplet with mass $m_x=\sqrt{2}g_x x_H\eta$, while the VEV of
$H$ generates a joint supersymmetric mass for $H^c$ and $S$
producing two chiral multiplets of mass $m_{\lambda}=\lambda\,\eta$.  The
fermion states of these multiplets mix to form a single Dirac
fermion (necessary on account of the accidental $U(1)$), while the
scalar states do not mix at all at tree-level.

  This simple picture is deformed by adding supersymmetry-breaking soft terms
to the potential.  The precise effect of the soft terms depends on
their origin, whether from gauge or gaugino mediation or due to
residual supergravity effects.  In general, however, the
phenomenology of the hidden sector can be classified according to
the scale of the hidden masses relative to the gravitino mass
$m_{3/2}$.  We again consider the cases $m_{3/2}\ll m_{hid}$ and
$m_{3/2} \gtrsim m_{hid}$.

\subsubsection{$\mathbf{m_{3/2}\ll m_{hid}}$}

When $m_{3/2}\ll m_{hid}$, little gauge or gaugino mediation
then provides the largest contributions to
the $U(1)_x$-sector soft terms, which are parametrically smaller
than the VEV induced by $\xi$, allowing us to again treat them as
small perturbations on the supersymmetric spectrum. The tree-level
scalar potential then becomes
\bea
\label{vnmssm}
V &=& (m_H^2-x_H g_x^2\xi)|H|^2 + \frac{x_H^2 g_x^2}{2}|H|^4\\
&&+ (m_{H^c}^2+x_H g_x^2\xi-g_x^2x_H^2|H|^2+|\lambda|^2|H|^2)|H^c|^2
+ \frac{x_H^2g_x^2}{2}|H^c|^4\nnmb \\
&&+ (m_S^2+|\lambda|^2|H|^2)|S|^2 + |\lambda|^2|S|^2|H^c|^2 -
(\lambda\,A_{\lambda}\,S\,H^c\,H + h.c.).\nnmb
\eea
With $x_H g_x^2 \xi >> m_{H^{(c)}}^2$, the VEV is now shifted to
\beq
\left<H\right>
\equiv \eta = \sqrt{\xi/x_H - m_{H}^2/(x_Hg_x)^2},
\eeq
while $\left<H^c\right> = 0 = \left<S\right>$ as before. As a result,
the potential maintains an exact $U(1)$ global symmetry under
which $H^c$ and $S$ are oppositely charged. The spectrum of states
still consists of a massive vector multiplet and a ``Dirac'' pair
of chiral multiplets, although the components within these
multiplets will now be split in their masses.

  Within the $\lambda$ sector of the theory derived from
$H^c$ and $S$, the bosonic states consist of a complex $h^c$
scalar derived from $H^c$ with mass
\beq m_{h^c}^2  =
m_{H^c}^2 + m_{H}^2 + |\lambda|^2\eta^2, \label{hcmass} \eeq as
well as a complex $h^s$ scalar derived from $S$ with mass \beq
m_{h^s}^2= m_s^2+|\lambda|^2\eta^2. \label{hsmass} \eeq There will
be a very small additional mixing between $h^c$ and $h^s$ on the
order of $A_{\lambda}\eta/m_H^2 \sim \sqrt{\epsilon}$ induced by
the subleading $A_{\lambda}$ term.  The unbroken accidental global
$U(1)$ in this sector ensures that the mixed mass eigenstates are
complex scalars with degenerate real and imaginary components.
This same $U(1)$ also ensures that the fermion state derived from
$S$ and $H^c$ is pure Dirac, and its tree level mass is still
$m_{\lambda} = \lambda\eta$.

  Among these states, the $h^s$ scalar will be the lightest
on account of the soft mass $m_s^2$.  This mass runs negative in
the IR due to the effect of the $\lambda$ coupling on the RG
evolution. In fact, neglecting subleading contributions to the
running from the $A$-terms, it is not hard to show that (to
one-loop order) we have $(m_{H}^2+m_{H^c}^2+m_s^2)\geq 0$,
$m_s^2\leq 0$, and $m_H^2=m_{H^c}^2\geq 0$ throughout the RG flow
to the IR, given little gauge or gaugino mediation boundary
conditions at the messenger or compactification scale.  The $h^s$
scalar is stable to the extent that the accidental global $U(1)$
is not accidentally broken.

  All states in the gauge sector have equal masses
up to small corrections.  The fermionic states retain a Dirac mass
of $m_x = \sqrt{2}x_H g_x\eta$ but acquire a tiny
$\epsilon$-suppressed Majorana splitting from the subleading
gaugino soft mass. Radiative effects also split these states apart
from the gauge boson by an amount on the order of
$\lambda^2\,m_s^2/16\pi^2$.  For example, the tree-level mass of
the physical real scalar $h$ is given by \beq m_h^2 =
{2}\,x_H^2g_x^2\,\eta^2 \label{hmass} \eeq as for the fermions and
the gauge boson, but is lifted by radiative corrections.  The
dominant effect is to raise the effective quartic coupling
$Q_{eff}$ analogously to the Higgs quartic in the MSSM.  This
produces the shift \beq Q_{eff} = 2\, x_H^2 g_x^2 \to 2\,x_H^2
g_x^2 + \frac{|\lambda|^4}{8\pi^2}\, \ln\left(\frac{
m_{h^c}^2\,m_{h^s}^2} {|\lambda|^4\eta^4}\right). \eeq The
logarithm is non-negative within little gauge or gaugino mediation
on account of the soft masses appearing in the masses of
$m_{h^c}^2$ and $m_{h^s}^2$ along with the properties of their RG
flow discussed above.  This pushes up the real scalar mass $m_h^2
= Q_{eff}\eta^2$ relative to the tree-level value. The other
states receive a radiative shifts in their masses of the same
order.  These splittings  are all very small relative to the
tree-level masses due to loop suppression as well as the hierarchy
$m_{soft} \sim \sqrt{\epsilon}\,\eta$.

  The phenomenology of this scenario depends primarily on which of
the two sectors is lighter.  Whether or not the $\lambda$ sector
is lighter, the unbroken global $U(1)$ ensures that the lightest
$h^s$ scalar is stable, while the slightly heavier Dirac fermion
state is metastable on account of $R$-parity (assuming all fields
are $R$-even) and annihilates efficiently into the lighter scalar.
When this chiral sector is lighter than the massive gauge sector,
the lightest $h^s$ scalar can only annihilate into lighter visible
sector particles through the $s$-channel exchange of a $U(1)_x$
gauge boson.  We estimate the cross-section for this process in
the early universe to be~\cite{Fayet:2004bw}
\bea \label{hssigma}
\left<\sigma v\right> &\simeq& \frac{2g_x^2x_H^2e^2c_W^2}{3\pi}
\frac{m_{h^s}^2}{(4\,m_{h_s}^2-m_{Z_x}^2)^2}
\,\epsilon^2\,|U_{cs}|^4\,v_{f.o.}^2\\
&\simeq&
(1\times 10^{-36}\,cm^3/s)\lrf{g_xx_H}{0.1}^2\lrf{m_{h_s}}{0.1\,\gev}^2
\lrf{1\,\gev}{m_{Z_x}}^4
\nnmb\\
&&~~\lrf{|U_{cs}|}{\sqrt{\epsilon}}^4\,
\lrf{\epsilon}{10^{-3}}^4\,\lrf{v_{f.o}}{0.3}^2,
\nnmb
\eea
where $v_{f.o.}$ is the typical particle velocity at thermal
freeze-out and $U_{cs} \sim \sqrt{\epsilon}$ denotes the small
mixing between the $h^s$ and $h^c$ states induced by
$A_{\lambda}\sim \epsilon^2M_1$. On account of this additional
mixing suppression, the singlet scalar relic density is overly
large.  This scenario is therefore unacceptable unless
higher-dimensional operators break the accidental $U(1)$ and allow
for an efficient decay of the singlet scalar~\cite{Katz:2009qq}.

  When the gauge sector is lighter, the lightest fermion in this sector
is metastable on account of $R$-parity.  It is nearly degenerate
with the $U(1)_x$ gauge boson and the real scalar $h$, and thermal
effects allow it to annihilate into gauge boson pairs.\footnote{
The radiative mass splitting is much smaller than the temperature
at thermal freeze-out.} The corresponding cross-section at
freeze-out is estimated to be \bea\label{fermann} \left<\sigma
v\right> &\simeq&
\frac{g_x^4x_H^4}{16\pi}\frac{1}{m_{Z_x}^2}\,v_{f.o.}\\
&\simeq& (7\times 10^{-24}
\mbox{cm}^3/\mbox{s})\lrf{g_x x_H}{0.1}^4\lrf{1\,\gev}{m_{Z_x}}^2
\lrf{v_{f.o.}}{0.3},\nnmb
\eea
where $v_{f.o.}$ denotes the typical relic velocity
during freeze-out.
This annihilation cross-section leads to a relic density smaller
than the measured value, though such a candidate may constitute a
subdominant component of the dark matter. The $h^s$ scalar will
also be stable in the present scenario. However, it is now able to
annihilate efficiently into the scalar state in the $U(1)_x$
sector, and its corresponding relic density will be tiny.

  Since we assume the gravitino is lighter than the gauge-sector LHP,
the fermionic relic will decay at a later time into a gravitino
and a photon.  This limit was considered in \cite{Cheung:2009qd}.
The corresponding lifetime is
\bea
\tau &=& \frac{16\pi\left<F\right>^2}{m_{x}^5\,|P_{\gamma}|^2}\\
&\simeq& (3\times 10^3\,s)\,
\lrf{\sqrt{\left<F\right>}}{100\,\tev}^4\lrf{1\,\gev}
{m_{x}}^5\,\lrf{\epsilon}{|P_{\gamma}|}^2\lrf{10^{-3}}{\epsilon}^2,
\nnmb
\eeq
where $P_{\gamma} \sim \epsilon$ is the projection of the light
$U(1)_x$-sector fermion onto the photino.
Energetic electromagnetic decays 
are strongly constrained by limits on photodissociation during
BBN. For the fermion relic density we estimate, Ref.~\cite{Kawasaki:2004qu}
indicates that this lifetime must be less than about
$\tau \lesssim 10^4\,s$. From this, we obtain the strong upper
bound on the scale of supersymmetry breaking, implying that this
scenario is only viable for extremely low gauge messenger scales.

\subsubsection{$\mathbf{m_{3/2}\gtrsim m_{hid}}$}

  Our findings for $m_{3/2}\ll m_{hid}$ lead us to
consider the opposite limit, $m_{3/2} \gtrsim m_{hid}$. As before,
we study the two cases of high-scale gauge mediation with
$M_{mess} \sim 10^{14}\,\gev$ or lower-scale gaugino mediation
with a compactification scale of about the same size.

  With high-scale gauge mediation without sequestering,
all soft terms will receive additional supergravity contributions
on the order of $m_{3/2}$.  Assuming $m_{3/2}\sim \gev$,
the MSSM superpartner spectrum is only slightly perturbed,
with any additional supergravity-induced flavor mixing being on the
limit of what is consistent with current flavor bounds.
The $U(1)_x$ sector, on the other hand, is significantly modified
and the precise spectrum depends on unknown UV physics.
Even so, we can identify a few general features.

  If the only field to condense is $H$, the theory will again
split into two subsectors.  The fermion mass matrix will take the
same form as Eq.~\eqref{fermionmassmatrix}, but now with a gaugino
soft mass of a similar size to the other mass terms.  This will
generate a gaugino-Higgsino state that is at least somewhat
lighter than the gauge boson, depending on the size of the soft
mass for the hidden gaugino.  If the gauge sector contains the
LHP, it will not have any hidden-sector annihilation modes, it
will only be able to annihilate through $s$-channel $U(1)_x$ gauge
boson exchange into the visible sector.  The annihilation
cross-section, assuming the gravity-mediated gaugino soft mass
splits the Dirac states sufficiently into two Majorana states, is
\bea \label{visann} \left<\sigma v\right> &\simeq&
\frac{g_x^2x_H^2e^2c_W^2}{12\pi}\,\epsilon^2 |U_{fh}|^4
\frac{m_{x}^2}
{(4m_{x}^2-m_{Z_x}^2)^2+\Gamma_{Z_x}^2m_{Z_x}^2}v_{f.o.}^2\\
&\simeq&  (2\times 10^{-29}\mbox{ cm}^3/\mbox{s})
\lrf{g_xx_H}{0.1}^2\lrf{\epsilon}{10^{-3}}^2|U_{fh}|^4\lrf{m_{x}}{1\,\gev}^2
\lrf{1\,\gev}{m_{Z_x}}^4\,\lrf{v_{f.o.}}{0.3}^2,\nnmb
\eea
where
$U_{fh}$ is the mixing between the Higgsino and the LHP.
This cross-section is somewhat too small for the sample parameters chosen,
but increases to an acceptable level for slightly lighter hidden-sector
masses, a larger $U(1)_x$ gauge coupling, or if there is a resonant
enhancement of the annihilation process.  Since the gravitino mass is now
of the same order as the hidden-sector masses, this state will either
be stable on account of $R$-parity or very long lived, and hence potentially
a good dark matter candidate.

  When the $\lambda$ sector is lighter and only $H$ condenses,
supergravity effects make a significant contribution to the soft
scalar masses for ${S}$ and $H^c$ as well as the $A_{\lambda}$
trilinear soft coupling.  This mixes the $\lambda$-sector scalars,
and the lighter of the two complex mass eigenstates can be heavier
or lighter than the Dirac fermion in this sector.  Both are
stable.  If the lightest scalar is the LHP, its relic abundance is
given by Eq.~\eqref{hssigma}, but now with $|U_{cs}|$ typically on
the order of unity due to the unsuppressed supergravity
contribution to $A_{\lambda}$.  This abundance can potentially be
acceptable for smaller masses, larger gauge couplings, or with a
resonant enhancement of the annihilation cross-section. The
heavier $\lambda$-sector fermion now has its relic abundance set
by annihilation to pairs of the LHP scalar with cross-section \bea
\label{lambdatohs} \left<\sigma v\right> &\simeq&
\frac{\lambda^4}{16\pi}\, |U_{x\lambda}|^4
\frac{1}{m_x^2}\\
&\simeq&  (2\times 10^{-23}\mbox{cm}^3/\mbox{s})
\lrf{\lambda}{0.1}^4 |U_{x\lambda}|^4 \lrf{1\,\gev}{m_x}^2,\nnmb
\eea where $m_x$ is the mass of the lightest gauge-sector fermion
and $U_{x\lambda}$ its coupling to the $\lambda$-sector fermion
and the LHP scalar.

  If the $\lambda$-sector Dirac fermion is the LHP,
its annihilation cross-section will be given by
Eq.~\eqref{visann}, but with the replacement $m_x \to m_{\lambda}$
and without the $p$-wave factor of $v_{f.o.}^2$. This can provide
an acceptable relic density for somewhat smaller masses, larger
couplings, or with a resonant enhancement of the annihilation. The
lightest $\lambda$-sector scalar $h_{\lambda}$ can now annihilate
efficiently into the Dirac fermion LHP. The corresponding
cross-section is approximately four times the value in
Eq.~\eqref{lambdatohs}, and the resulting scalar relic density is
expected to be safely small.

  If the $H$, $H^c$, and $S$ fields all condense due to additional
supergravity contributions to the soft masses, the global $U(1)$ in
the $\lambda$ sector is broken spontaneously leading to a massless boson.
This can be avoided by including a $\kappa\,S^3/3$ coupling to
the superpotential, or if there is explicit breaking by supergravity
effects.  The precise spectrum in this case will be UV dependent,
but by analogy with the usual NMSSM scenario, it is possible to make
all states in the hidden sector heavier than the $U(1)_x$ gauge boson.
In this case, the lightest stable state in the $U(1)_x$ sector
will annihilate efficiently to gauge bosons, and will typically
have a safely small relic density.  There is also the possibility
of having a stable LHP due to $R$-parity which is mostly singlet,
whose relic density can account for the observed dark matter abundance.

  When supergravity effects are sequestered, as in gaugino mediation,
additional contributions to the soft masses still arise from
anomaly mediation, but they are suppressed by a loop factor
relative to $m_{3/2}$.  For $m_{3/2} \sim m_{hid}$ in this
situation, we obtain an acceptable phenomenology provided $g_x <
\lambda$.  As above, this implies that the lightest $R$-odd hidden
particle will be a gaugino-Higgsino mixture that is approximately
degenerate with the hidden gauge boson.  Since there is an
approximate degeneracy, the fermion can still have phase space
available to annihilate into hidden gauge bosons and obtain an
acceptable relic density as in Eq.~\ref{fermann}, even when the
cross-section for annihilation into the visible sector in
Eq.~\ref{visann} is small.  Since the gravitino can be heavier
than the lightest fermion, there is again no problem with BBN
constraints from late decays into a gravitino and a photon, and
this state can again be stable and a component of the dark matter.
As in the case of the bare $\mu'$ model, we have found $m_{3/2}
\gtrsim m_{hid}$ is desirable for a viable phenomenology.

  Sequestering of supergravity effects also allows
for $m_{3/2} \gg m_{hid}$ such that the anomaly-mediated soft-mass
contributions are on the order of $\Delta m_{hid} \sim
m_{3/2}\,g_x^2/(4\pi)^2$. This is precisely the situation
considered in Ref.~\cite{Katz:2009qq}.  As in the case of
high-scale gauge mediation with $m_{3/2}\sim m_{hid}$, this leads
to a significant change in the $U(1)_x$-sector spectrum (which is
now largely calculable), with a similar effect. The LHP is either
a Majorana Higgsino-gaugino mixture, or comes from the $\lambda$
sector of the theory, and in both cases leads to too large of a
relic density.  This can be avoided by adding additional operators
allowing for a rapid decay of this state to the visible sector, as
suggested in Ref.~\cite{Katz:2009qq}.

\subsection{Singlet-Mediated SUSY Breaking\label{singletmed}}

  We now consider a scenario where a singlet field communicates
between the supersymmetry-breaking sector and the $U(1)_x$ hidden
sector.  This gives rise to several new features which we
investigate below. The superpotential is taken to be
\begin{equation}
W \supset \lambda S H H^c +\frac{\kappa}{3}S^3.
\label{ssuperp}
\end{equation}
The new ingredient is that we now assume that the $S$ field
couples directly to the supersymmetry-breaking sector, generating
a \emph{positive} scalar soft mass
\begin{equation}
m_{S}^2 \simeq (100 \mbox{ GeV})^2.
\end{equation}
Soft terms $A_{\lambda}$ and $A_{\kappa}$ corresponding
to the interaction of Eq.~\eqref{ssuperp} may also be generated,
but will be suppressed if $S$ couples primarily to a $D$-term
($R$-preserving) source of supersymmetry breaking.
The coupling of $S$ to $H$ and $H^c$ communicates supersymmetry
breaking to the hidden sector, generating soft masses of size
\begin{equation}
m_{H}^2 = m_{H^c}^2 \simeq -\frac{2\lambda^2}{16 \pi^2} m_{S}^2
\ln\lrf{\Lambda}{m_{hid}},
\end{equation}
where $\Lambda$ is the scale at which the soft mass $m_S^2$ is
generated. These hidden-sector soft masses are on the order of a
few GeV provided the coupling $\lambda$ is somewhat small and the
logarithm is not too large.

  The resulting scalar potential is identical to that given
in Eq.~\eqref{vnmssm}, but with additional terms proportional to $\kappa$.
With negative soft masses for $H$ and $H^c$,
both $H$ and $H^c$ can develop VEVs, breaking the $U(1)_x$ gauge
symmetry and giving the $U(1)_x$ gauge boson
a mass~\cite{Hooper:2008im,Zurek:2008qg}.
To analyze the mass spectrum of this scenario, we make the simplifying
assumptions that the kinetic mixing is suppressed, that $A_{\lambda}$
and $A_{\kappa}$ are somewhat suppressed relative to $m_S^2$,
and that $\kappa \ll 1$.  Under these assumptions, and assuming further
that $m_S^2 > 0$ at the low scale, we obtain
\beq
\langle H \rangle^2 \simeq \langle H^c \rangle^2 \equiv \eta^2
\simeq -\frac{m_{H}^2}{\lambda^2},
\eeq
as well as
\beq
\left<S\right> \equiv s =
\frac{\lambda A_{\lambda}\eta^2}{m_S^2}\simeq
-\frac{\lambda A_{\lambda}m_H^2}{\lambda^2m_S^2}.
\eeq
This minimum is stable provided $g_x^4-(\lambda^2-g_x^2)^2>0$.
This yields typical values of $\eta$ in the range of 20--60\,GeV
and somewhat smaller values of $s$.

  The Higgs VEVs break the gauge symmetry, giving the $U(1)_x$
gauge field a mass
\begin{equation}
m_{Z_x} = \sqrt{2}\,g_xx_H \eta.
\end{equation}
Expanding the scalars around their VEVs and removing the Goldstone
state eaten by the $U(1)_x$ gauge boson, $H^{(c)} = \eta +
({h^{(c)}+iA_H/\sqrt{2})/\sqrt{2}}$ and $S = s +
(h_s+iA_S)/\sqrt{2}$, the approximate mass eigenstates of the
CP-even scalars are given by $h_1\equiv
\frac{1}{\sqrt{2}}\left(h-h^c\right)$, $h_2\equiv
\frac{1}{\sqrt{2}}\left(h+h^c\right)$, and $h_s = h_s$ with approximate
masses
\bea
m_{h_1}^2 &\simeq& (4 x_H^2 g_x^2-2\lambda^2)\eta^2\\
m_{h_2}^2 &\simeq& 2\lambda^2\eta^2\\
m_{h_s}^2 &\simeq& m_S^2. \eea For the CP-odd scalar masses we
must take more care. In the limits of $\kappa \to 0$ or
$A_{\lambda,\kappa}\to 0$, the theory has a global Abelian
symmetry that is spontaneously broken by the VEVs, leading to a
massless Nambu-Goldstone boson, as can be seen from the axion mass
matrix in the $(A_H,\,A_S)$ basis
 \begin{equation}
{\cal M}_A^2 = \left(
\begin{array}{cc}
2\lambda\,A_\lambda s - 2\lambda \kappa s^2 &
\sqrt{2}\lambda A_{\lambda}\eta +2\sqrt{2}\lambda\kappa\eta s\\
\sqrt{2}\lambda A_{\lambda}\eta +2\sqrt{2}\lambda\kappa\eta s&
\lambda A_{\lambda}\frac{\eta^2}{s}+3\kappa A_{\kappa}s-4\lambda\kappa\,\eta^2
\end{array}
\right).
\label{axionmass}
\end{equation}
Under our assumption that these parameters are relatively small,
there remains a light pseudo-axion in the spectrum with mass
\beq
m_{a_1}^2 \simeq 6\frac{s^2}{\eta^2}\,
(-3\lambda\kappa\eta^2+\kappa A_{\kappa}s).
\eeq
This state derives mostly from $H$ and $H^c$, and the expression
for its mass implies that $\lambda\kappa < 0$ is needed for the
stability of the perturbed minimum.
The second pseudoscalar is mostly singlet and has mass
$m_{a_2}^2\simeq m_S^2$.

  The scalar VEVs also induce a mixing between the hidden-sector
gauginos and Higgsinos.  In the basis
$(\tilde{\lambda},\tilde{H},\tilde{H}^c,\tilde{S})$ the fermion
mass matrix is \beq \mathcal{M}^f = \left(
\begin{array}{cccc}
0&\sqrt{2}x_H g_x\eta&-\sqrt{2}x_H g_x\eta&0\\
\sqrt{2}x_H g_x\eta&0&0&\lambda\,\eta\\
-\sqrt{2}x_H g_x\eta&0&0&\lambda\,\eta\\
0&\lambda\,\eta&\lambda\,\eta&0
\end{array}
\right).
\eeq
This gives two fermions with mass $M^f_{1,2} = 2 x_Hg_x \eta$
and two fermions with mass $M^f_{3,4} = \sqrt{2} \lambda \eta$.

  With our simplifying assumption of small
$A_{\lambda}$ and $A_{\kappa}$, the lightest state in the hidden
$U(1)_x$ sector is the pseudo-axion. In principle, these $A$-terms
can be the same order as $m_S$ if the singlet receives its soft
mass through couplings to an $R$-breaking source of supersymmetry
breaking, such as the gauge messengers. On the other hand, these
terms can be significantly smaller than $m_S^2$ if the dynamics
generating $m_S^2$ preserves an R-symmetry. In this case the
$A$-terms will be set by the dominant contribution to R-breaking
in the hidden sector, which could arise from supergravity or
renormalization group effects, and the lightest hidden-sector
state will be the light pseudo-axion.  Depending on the mass of
the gravitino, the lightest $U(1)_x$ fermion will be stable or
metastable on account of $R$-parity.\footnote{Preservation of
$R$-parity requires that $S$, $H$, and $H^c$ are all even.} Thus,
the phenomenology of the hidden sector depends on the gravitino
mass, so we again consider the two cases $m_{3/2} \ll m_{hid}$ and
$m_{3/2} \gtrsim m_{hid}$.

\subsubsection{$\mathbf{m_{3/2} \ll m_{hid}}$}

  With $m_{3/2} \ll m_{hid}$, the lightest $U(1)_x$-sector
fermion will be unstable against decaying to a pseudo-axion
and a gravitino.  The lifetime for this decay is
\bea
\tau &=& \frac{16\pi\left<F\right>^2}{m_{\chi}^5\,|P_{\chi\tilde{a}}|^2}\\
&\simeq& (3\times 10^{-3}\,s)\,
\lrf{\sqrt{\left<F\right>}}{100\,\tev}^4\lrf{1\,\gev} {m_{\chi}}^5
\frac{1}{|P_{\chi\tilde{a}}|^2}, \eeq where $P_{\chi\tilde{a}}$ is
the projection of the lightest fermion onto the superpartner of
the pseudo-axion.  This decay will occur safely before
nucleosynthesis provided the $F$-term parameterizing supersymmetry
breaking is not too large.

  The light pseudo-axion in the hidden sector will be stable in the
absence of any interactions with the SM.  If these axions are
heavier than an eV and completely stable, they create a problem
with the cosmological abundance.  On the other hand, they will
decay away efficiently if they couple even very weakly to the SM.
For example, if $S$ has a small coupling to the visible sector of
the form $\zeta S H_u H_d$, the pseudo-axion will mix with the
CP-odd Higgs and decay to photon
pairs~\cite{Kalyniak:1985ct,Gunion:1988mf}.  The leading coupling
in this case comes from the cross-term in the $F$-term potential
due to $S$, and leads to a pseudo-axion lifetime on the order of
\bea \label{axtau} \tau &\simeq&
\frac{256\pi^3}{\alpha^2\lambda^2\zeta^2}\,
\frac{m_A^4}{\eta^2m_{a_1}^3}\\
&\simeq&
(0.6\,s)\,\lrf{10^{-3}}{\zeta}^2\lrf{0.1}{\lambda}^2
\lrf{m_A}{100\,\gev}^4
\lrf{40\,\gev}{\eta}^2
\lrf{0.1\,\gev}{m_{a_1}}^3.
\nnmb
\nnmb
\eea where $m_{A}$ is the mass of the MSSM pseudoscalar. This
lifetime can thus be made safe in terms of cosmology, but is very slow
relative to particle-collider timescales.

\subsubsection{$\mathbf{m_{3/2} \gtrsim m_{hid}}$}

In the case that $m_{3/2} \gtrsim m_{hid}$, the lightest fermion
in the $U(1)_x$ sector will typically be the LSP, and is stable on
account of $R$-parity.  This state can annihilate efficiently
into pseudo-axions.  When the $\lambda < \sqrt{2}\,g_x$, the annihilation
cross-section is on the order of
\beq
\langle \sigma v \rangle  & = &
\frac{\lambda^4}{4\pi}|U_{hx}|^4 \frac{1}{(\sqrt{2}\lambda \eta)^2}  \\
&\simeq&  (1 \times 10^{-23} \mbox{ cm}^3/\mbox{s})
\left(\frac{\lambda}{0.1}\right)^4 |U_{hx}|^4\left(\frac{3~ {\rm
GeV}}{\sqrt{2}\lambda\eta}\right)^2, \nonumber \eeq where $U_{hx}$
is a gaugino-Higgsino mixing factor close to unity. We obtain a
similar cross-section with $\lambda \to \sqrt{2}g_xx_H$ when
$\lambda > \sqrt{2}g_xx_H$. This cross-section is large enough
that the relic abundance of the stable fermion is safely small,
though this state may compose a fraction of the dark matter.

\subsection{Multi-Mediator Models\label{NMSSMGauge}}

  A minimal model with bi-fundamental mediators to
a hidden $U(1)_x$ sector consists of a pair of chiral
bi-fundamentals $F$ and $F^c$ that transform as $Y=\pm 1/2$
doublets under $SU(2)_L$ and have charges $\pm x_F$ under
$U(1)_x$.\footnote{ To preserve MSSM gauge unification, we could
also incorporate $F$ and $F^c$ into a set of
$\mathbf{5}\oplus\mathbf{\overline{5}}$'s.} In addition to these
states, we assume that there also exists a set of fields charged only
under $U(1)_x$, such as the models described above.  For
simplicity, we take the superpotential for the bi-fundamentals to be
\beq
W\supset \mu_FF\,F^c,
\eeq
along with the soft-breaking operator
\beq
V_{soft} \supset -\left[(B\mu)_F F\,F^c+h.c.\right].
\eeq
While we do not specify the origin of these terms, which we assume
to be of $\tev$ size, they could originate dynamically from an
NMSSM-like mechanism.  Conversely, if the mass term $\mu_F$ is
supersymmetric in origin, the corresponding $(B\mu)_F$ soft operator
will arise from RG running.  Pushing $\mu_F$ to be larger
than the gauge messenger scale with $(B\mu)_F\to 0$, we could integrate
these states out and their effect would be felt only through their
contribution to kinetic mixing.

    Bi-fundamental mediators can be added to any of the light $U(1)_x$
sectors described previously.  Their effect on the properties of
the light sector depends importantly on the size of the $U(1)_x$
gaugino mass they induce.  As discussed in
Section~\ref{addmedfields}, if $\mu_F$ is a genuinely
supersymmetric threshold, gaugino screening will occur and the
$U(1)_x$ gaugino mass will arise only at 5-loop 
order~\cite{ArkaniHamed:1998kj}.  On the other hand, if the 
bi-fundamental threshold is not completely
supersymmetric, the contribution to the $U(1)_x$ gaugino mass
will be on the order of
$(g_x^2x_F^2/(4\pi)^2)(B\mu)_F/\mu_F$~\cite{Poppitz:1996xw}.

  In the screened case, mediator effects will only significantly
modify the soft masses of the light hidden-sector scalar fields.
For smaller $g_x$, the shifts in the soft masses are subleading
relative to the effect of the induced FI term, and the
phenomenology of the light states will remain similar to that
described above.\footnote{With lighter bi-fundamental mediators
the gauge coupling must be relatively small, $g_x \lesssim 0.1$,
to avoid generating an unacceptably large low-scale value for the
kinetic mixing $\epsilon$ through RG effects, Eq.~\eqref{epsilonloop}.}
In particular, the $\mu'$ model with
GeV-scale residual supergravity contributions as well as the
NMSSM scenario with a lighter gauge sector with either small
$m_{3/2}$ or $m_{3/2}\sim \gev$ are viable scenarios.

  When the bi-fundamental mass threshold is not supersymmetric,
the mediators will contribute significantly to both the gaugino
and scalar soft masses in the light $U(1)_x$ sector.
Here, both the $\mu'$ model with residual supergravity effects
and the NMSSM model with small $m_{3/2}$ can be phenomenologically
acceptable.  However, the NMSSM model with larger $m_{3/2}$
(and only $H$ condensing) will have a problematic gaugino-Higgsino state
lighter than the $U(1)_x$ gauge boson, on account of the mediator
contribution to the gaugino mass.  This tends to produce
too large of a relic density unless there is an enhancement
of the annihilation cross-section relative to the estimate
in Eq.~\eqref{visann}.

\section{Signatures in Dark Matter Searches and Colliders\label{apps}}

  In this section we consider using the GeV-scale hidden sectors
studied above to help provide dark matter explanations
for some of the intriguing signals seen in DAMA, PAMELA, ATIC, and
PPB-BETS.  We also discuss how these light hidden sectors might be probed
at present and future particle-collider experiments.

\subsection{Dark Matter Direct Detection and DAMA}

  The DAMA/NaI and DAMA/LIBRA experiments, consisting of NaI-based
scintillation detectors, have reported an annual modulation signal
with a significance of $8.3\sigma$~\cite{Bernabei:2008yi}.  Both
the period and the phase of this modulation are consistent with
dark matter scattering off detector nuclei.  The main challenge of
such a dark matter interpretation, however, is to maintain
consistency with the null results of other dark matter direct
detection searches, such as CDMS~\cite{Ahmed:2008eu} and
XENON~\cite{Angle:2007uj}.  Two possibilities that are potentially
consistent with both DAMA and other null result bounds are light
($m\lesssim 10\,\gev$) elastically-scattering dark
matter~\cite{Gelmini:2004gm,Petriello:2008jj,
Savage:2008er,Savage:2009mk,Andreas:2008xy,Kim:2009ke}, 
and heavier ($m\gtrsim 50\,\gev$)
inelastically-scattering dark
matter~\cite{TuckerSmith:2001hy,TuckerSmith:2002af,TuckerSmith:2004jv,Chang:2008gd,MarchRussell:2008dy,Cui:2009xq}.

\subsubsection{Elastic Dark Matter}

  Light elastic dark matter can produce observable recoils at DAMA,
while remaining consistent with other direct detection null
results. This occurs in a window where the dark matter has a mass
between about 3 and 10 GeV and a spin-independent scattering
cross-section in the range $\left(10^{-41}-10^{-39}\right)
\,\mbox{cm}^2$, and depends importantly on the phenomenon of
\emph{channeling}~\cite{Bernabei:2008yi,Petriello:2008jj,Savage:2008er}.\footnote{
There is also a spin-dependent scattering
window~\cite{Ullio:2000bv,Savage:2004fn,Savage:2008er}, but it has
been essentially closed by Super-Kamiokande constraints on
annihilating dark matter in the
sun~\cite{Hooper:2008cf,Feng:2008qn}. There is an exception,
however, if the dark matter is not self-annihilating, in which
case the constraints vanish.} The allowed window is constrained by
the spectral shapes of the DAMA modulated and unmodulated signal
rates~\cite{Chang:2008xa,Fairbairn:2008gz}, but there remains an
allowed region even after constraints from the spectrum of the signal are
taken into account~\cite{Savage:2009mk}.

  The models constructed above often contain a stable state
in the multi-GeV mass range with a thermal relic density close to
the measured dark matter value.  Such a state could potentially
act as a light elastic dark matter candidate, making up either the
majority or a significant fraction of the total relic density of
dark matter, provided it has an acceptable nucleon scattering
cross-section.  The simplest scattering mechanism in these
scenarios consists of nuclear scattering mediated by the light
$U(1)_x$ gauge boson. This state can effectively mix with electric
charge through kinetic mixing, and in the gauge diagonal basis the
visible sector states acquire $U(1)_x$ charges equal to
$-eQc_W\epsilon/g_x$, where $Q$ denotes the electric charge of
that state.  Consequently a potential dark matter state with
$U(1)_x$ charge $x_{DM}$ has an effective nuclear-scattering
cross-section off of a proton equal to~\cite{Jungman:1995df}
\beq
\sigma_p \simeq \frac{\mu_p^2}{\pi}\lrf{g_x}{M_{Z_x}}^4\,
\left(\frac{e\,c_W\epsilon}{g_x}\right)^2\,x_{DM}^2,
\label{sicx}
\eeq
where $\mu_p \simeq m_p$ is the reduced mass of the
proton-DM system.

  When the $U(1)_x$ breaking is dominated by the hypercharge FI term
induced by the visible-sector Higgs VEVs, we can further reduce
this expression.  In this case, we obtain
a gauge boson mass of $m_{Z_x}^2 \simeq
{g_xg_Y}{|c_{2\beta}\,x_H|}\epsilon\,v^2.$  Plugging this into
Eq.~\eqref{sicx}, the factors of $\epsilon$ and $g_x$ amazingly
cancel out, and we find \beq \sigma_p \simeq (5\times 10^{-38}
\mbox{ cm}^2)\,\lrf{\mu_p}{m_p}^2\,\frac{1}{c_{2\beta}^2}\,
\lrf{x_{DM}}{x_H}^2. \eeq Unless there is a hierarchy in
$x_{DM}/x_H$, this cross-section is too large (assuming the local
dark matter density is $0.3\,\gev/\mbox{cm}^3$) by about two orders of
magnitude. It can, however, be reduced if there are additional
contributions to $\xi_Y$, or if the $U(1)_x$ symmetry breaking is
driven by other soft parameters, as in the singlet-mediated model.

 A second possible scattering mechanism for a light DM candidate
arises if there is a hidden-sector singlet which has a small
coupling to visible-sector fields, such as through the terms
$\lambda S H H^c + \zeta S H_u H_d$.  Such couplings induce a small
mixing among the visible- and hidden-sector Higgs states.
For the models considered in Sections~\ref{hidsecnmssm}
and \ref{singletmed}, the scattering cross-section
of the lightest $U(1)_x$ sector fermion off a nucleon
is approximately
\bea
\sigma_n &\simeq& \frac{\mu_n^2}{\pi}N_n^2|U|^4\,
\lrf{\lambda\zeta\,v_u\eta}{m_{h^0}^2}^2\frac{1}{m_{h_1}^4}\\
&\simeq& (2\times 10^{-41}\,cm^2)\,
\lrf{\mu_n}{m_p}^2\lrf{N_n}{0.1}^2|U|^4\lrf{\lambda}{0.1}^2
\lrf{\zeta}{10^{-3}}^2\nnmb\\
&&~~\lrf{\eta}{20\,\gev}^2\lrf{115\,\gev}{m_{h^0}^2}^4
\lrf{3\,\gev}{m_{h_1}}^4,\nnmb \eea where $N_n$ comes from the
effective coupling of the exchanged scalar to the target nucleus,
$U$ corresponds a mixing factor of order unity, $h^0$ is the MSSM
Higgs, and $h_1$ is a hidden-sector $CP$-even Higgs scalar. This
cross-section is somewhat small for light elastic DM, but may be
enhanced for larger values of $\zeta$ or smaller values of the
hidden-sector scale.

\subsubsection{Inelastic Dark Matter}

  A second potential dark matter explanation for the DAMA annual
modulation signal is inelastic dark
matter~(IDM)~\cite{TuckerSmith:2001hy,TuckerSmith:2002af,TuckerSmith:2004jv,
Chang:2008gd,MarchRussell:2008dy,Cui:2009xq}. In contrast to
elastic-scattering dark matter, IDM scatters preferentially off
target nuclei into a second slightly heavier state.  This enhances
the annual modulation of the signal and modifies the kinematics of
the scattering process such that the scattering rate off heavier
nuclear targets, such as the iodine in DAMA, is enhanced relative
to lighter elements, such as the germanium used in CDMS.
Inelastic dark matter can then account for the DAMA signal while
being consistent with other direct detection bounds for a wide
range of dark matter masses (above about $50\,\gev$).  This
requires an inelastic mass splitting on the order of $100\,\kev$
and an effective nucleon scattering cross-section in the range
$\sigma_n \sim \left(10^{-40}-10^{-38}\right)\,\mbox{cm}^2$,
assuming a single dominant dark matter
component~\cite{TuckerSmith:2001hy,
TuckerSmith:2002af,TuckerSmith:2004jv,Chang:2008gd,
MarchRussell:2008dy,Cui:2009xq}.

  IDM can arise naturally from a Dirac fermion or a complex scalar
whose real components are split slightly in mass, and that couples
to nuclei primarily through a massive gauge
boson~\cite{TuckerSmith:2001hy}.  The couplings of the resulting
mass eigenstates to the gauge boson then connect states with
different masses, naturally giving rise to an inelastic
interaction.  Among the possibilities for the massive gauge boson
mediating nuclear scattering is a light hidden $U(1)_x$ that
couples to the visible sector only through gauge kinetic mixing.
The effective scattering cross-section for a dark matter particle
of charge $x_{DM}$ off a proton mediated by such a gauge boson was
estimated in Eq.~\eqref{sicx}. As for light elastic DM, this
cross-section is slightly too large when the hidden sector
symmetry breaking is dominated by the induced hypercharge FI term
unless there is a hierarchy between $x_{DM}$ and $x_H$ or if there
are other contributions to the gauge boson mass. On the other
hand, such a large nucleon scattering cross-section can be
acceptable if the IDM makes up only a small fraction of the total
dark matter density.

  Coupling a TeV-scale dark matter state to a GeV-scale hidden
sector can induce an inelastic mass splitting of the right
size~\cite{
ArkaniHamed:2008qn,Baumgart:2009tn,Cui:2009xq,Cheung:2009qd,Katz:2009qq,
Alves:2009nf}. Consider introducing a vector pair of chiral states
$D$ and $D^c$ charged under $U(1)_x$, and coupling them to the
condensing Higgs $H$ in the hidden sector as well as a pair of
chiral singlets $N_1$ and $N_2$, \beq W \supset
\xi_DN_1\,DD^c+\xi_N N_1N_2^2 + \zeta\,D H N_2. \eeq If $N_1\to
\left<N_1\right> \sim \tev$, we can integrate out the $N_2$ state
to get the effective superpotential \beq W_{eff} \supset
-\frac{\zeta^2}{2\xi_N\left<N_1\right>}(D H)^2. \eeq This operator
yields an inelastic mass splitting that is naturally on the order
of a few hundred $\kev$ through the numerology $\mev \sim
\gev^2/\tev$.  Note, however, that this operator requires $x_H =
x_{DM}$, which implies a nucleon scattering cross-section that is
too large for acceptable IDM making up the full relic density when
the symmetry breaking in the hidden sector is dominated by the
hypercharge FI term induced by the Higgs VEVs.

Unfortunately, the minimal IDM model presented above is
problematic because the heavier inelastic state tends to be very
long-lived, and typically develops an unacceptably large relic
density~\cite{Finkbeiner:2009mi,Batell:2009vb}.  This difficulty
can be avoided if the $D$ and $D^c$ states carry SM
charges in addition to the $U(1)_x$, as the heavier inelastic
state will now be able to decay to the lighter state and neutrinos
through a $Z^0$ gauge boson.  Furthermore, the cross-section
obtained from scattering through $Z^0$ exchange is roughly the
correct size to account for the DAMA signal.

  The simplest way to realize this scenario is to take
$D$ and $D^c$ to be $SU(2)$ doublets with hypercharge $Y=\mp 1/2$.
An inelastic mass splitting can then be generated in a number of ways.
Introducing a pair of states $X_1$ and $X_2$ with $U(1)_x$ charges $\pm x_H/2
= \mp x_{D}$, we can write
\beq
W \supset \lambda_1DH_uX_1 + \frac{\lambda_2}{2}HX_2^2 + M_xX_1X_2,
\eeq
which, after supersymmetrically integrating out $X_1$ and $X_2$,
generates the effective superpotential
\beq
W_{eff} \supset \frac{\lambda_1^2\lambda_2}{2M_x^2}H(D H_u)^2.
\eeq
This operator can generate the correct
inelastic splitting for $\lambda_1\sim \lambda_2 \sim 0.1$ and
$M_x \sim \tev$ when $\left<H\right> \sim \gev$.

  A second related scenario that is able to induce the correct inelastic
splitting consists of a singlet $S$ in addition to $X_1$ and
$X_2$, and the superpotential \beq W \supset \lambda_1 D H_uX_1 +
\lambda_2H X_2S + \frac{1}{2}M_sS^2 + M_xX_1X_2, \eeq where $X_1$
and $X_2$ now have $U(1)_x$ charges $\pm x_H$. Integrating out
$N$, $X_1$, and $X_2$ at the supersymmetric level then generates
the inelastic mass-splitting operator \beq W_{eff} \supset
-\frac{\lambda_1^2\lambda_2^2}{2M_sM_x^2}(DH_uH)^2. \eeq In this
case, we obtain an inelastic mass splitting on the order of
$100\,\kev$ for $M_x\sim M_s \sim 300\,\gev$, $\left<H\right> \sim
2\,\gev$, and $\lambda_1\sim \lambda_2 \sim 0.5$.

  The two IDM scenarios with bi-fundamental $D$ and $D^c$ fields
described above are very similar to the multi-mediator models
discussed in Section~\ref{NMSSMGauge}.  We found that these models
can lead to a phenomenologically acceptable GeV-scale $U(1)_x$
sector in a number of ways.  Scattering of the bi-fundamental IDM
off nuclei will be mediated both by the SM $Z^0$ as well as the
light $U(1)_x$ gauge boson (provided there is kinetic mixing).
Somewhat amusingly, the scattering of this IDM state off protons
will be dominated by the $U(1)_x$ gauge boson exchange, while its
scattering off neutrons will be dominated by the SM $Z^0$. The
relic density of the IDM will be determined in a large part by its
annihilation to gauge bosons.  For fermionic $SU(2)_L$-doublet
IDM, this implies that the mass of the state must be greater than
about $1000\,\gev$ to provide the observed relic density, while
for scalars the mass should be in excess of about
$500\,\gev$~\cite{Cirelli:2005uq}.  However, lighter IDM that
makes up only a small fraction of the total dark matter relic
density can still potentially account for the DAMA signal owing to
the often large proton scattering cross-section mediated by the
$U(1)_x$ gauge boson.

\subsection{Applications to PAMELA and ATIC}

  The PAMELA~\cite{Adriani:2008zr}, ATIC~\cite{Chang:2008zz},
and PPB-BETS~\cite{Torii:2008xu} experiments observe excesses
in cosmic ray positrons and electrons at energies above
$10\,\gev$.  These signals could potentially originate from dark matter
annihilating in our galaxy.  For such an explanation to work,
the dark matter state must be heavier than about $100\,\gev$ and annihilate
efficiently into leptons with a cross-section larger than the value
providing the correct thermal relic density~\cite{Cirelli:2008pk}.

  The light $U(1)_x$ models outlined above, when coupled
to a heavier dark matter state charged under the $U(1)_x$, can
have the correct properties to induce the necessary enhancement in
a subset of the phenomenologically consistent parameter
space~\cite{ArkaniHamed:2008qn}.  Annihilation of the heavy
$U(1)_x$-charged dark matter state into $U(1)_x$ gauge bosons in
our local region of the galaxy will generically receive a
Sommerfeld enhancement provided $\alpha_xm_{DM}/m_{Z_x}\gtrsim 1$.
To account for the signals at PAMELA or ATIC without violating
observational constraints on fluxes of gamma
rays~\cite{Bertone:2008xr,Bergstrom:2008ag,
Meade:2009rb,Mardon:2009rc} and
anti-protons~\cite{Cirelli:2008pk}, these dark gauge bosons should
subsequently decay primarily to leptons.  This occurs
automatically due to kinematics for $m_{Z_x}\lesssim 0.3\,\gev$
provided the $U(1)_x$ gauge boson is also lighter than twice the
mass of any of the other hidden-sector states.

\subsection{Collider Phenomenology}

  The presence of supersymmetric hidden sectors at the GeV
scale can lead to a variety of interesting signals at particle
colliders. First, at very high-energy colliders such as the
Tevatron and the LHC, the visible-sector LSP produced in cascade
decays is unstable against subsequently decaying into
hidden-sector states, as discussed in
Ref.~\cite{Strassler:2006qa}, and more recently in
Refs.~\cite{Hooper:2008im,ArkaniHamed:2008qn,Zurek:2008qg,Baumgart:2009tn}.
The hidden-sector particles may then cascade further, potentially
giving rise to additional visible- and hidden-sector final states.
Second, at high luminosity $e^+ e^-$ machines, such as the B and
charm factories, heavy-flavor mesons will have rare exotic decays
into the hidden sector, such as $e^+e^- \rightarrow \Upsilon
\rightarrow \gamma + hidden$, and $e^+ e^- \rightarrow \gamma +
hidden$ through an ISR photon.  For the models discussed in the
previous sections, we find that their collider signatures involve
photons plus missing energy, and in some cases highly collimated
``lepton jets''~\cite{ArkaniHamed:2008qn,Baumgart:2009tn}.  We
give a brief overview of potential collider signatures here,
leaving a more detailed study for future work.

\subsubsection{High-Energy Hadron Colliders}

  We consider first some of the potential signatures of an Abelian hidden
sector at high-energy hadron colliders such as the Tevatron and the LHC.
The production of hidden-sector particles will arise primarily
through the cascade decays of heavier states carrying SM charges.
In particular, the visible-sector LSP will itself be produced through
cascade decays in the usual way, and may subsequently decay into the
lighter hidden sector.  Hidden-sector states can also be produced
through the decays of (necessarily) heavy bi-fundamental states
charged under both the SM and $U(1)_x$ gauge groups.
This is a very simple example of a hidden valley
scenario~\cite{Strassler:2006im}.
For both cases, additional visible-sector states may also be
emitted in the LSP or bi-fundamental decay.

  The subsequent cascade in the hidden sector
can be a source of further visible-sector particles. The general
condition for this is that $U(1)_x$ gauge boson decay
predominantly into the visible sector, which usually requires that
it is kinematically incapable of decaying to pairs of
hidden-sector particles.  Other hidden-sector states may also have
decay modes to the visible sector, but they are generally very
slow on collider time scales in the scenarios considered above.
When the $U(1)_x$ gauge boson decays primarily to hidden states,
the hidden-sector cascades will unfortunately remain hidden.

  Consider the case where the visible-sector LSP (vLSP) is a neutralino,
and the $U(1)_x$ gauge boson decays mainly back to visible states.
The decay chains with a squark or slepton vLSP will be similar,
but with one more quark or lepton.
The dominant decay mode of the neutralino is $\chi^0 \rightarrow H
\tilde{H}$, where $H$ and $\tilde{H}$ are
hidden Higgs and Higgsino fields~\cite{Baumgart:2009tn}.  These states
then cascade down to the lightest fermion and scalar fields
in the hidden sector allowed by phase space.
Along the way, one or more $U(1)_x$ gauge bosons can be emitted.
These will typically decay promptly to highly collimated leptons
(and possibly pions), giving rise to low-invariant mass
``lepton jets''~\cite{Han:2007ae,ArkaniHamed:2008qn,
Baumgart:2009tn,Bai:2009it}.
In the models we have considered, both the final state fermions
and scalars (including the R-even states) are long-lived on collider
scales and leave the detector (see Eqs.~(\ref{hdecay4e},~\ref{axtau})).
This necessarily gives rise to a large component of missing energy
accompanying the ``lepton jet''.  Note, however, that the non-Abelian
models considered in \cite{Han:2007ae,ArkaniHamed:2008qn} have
characteristically busy events due to showering in the hidden sector
and the final-state leptons will tend to be somewhat soft.  On the other
hand, the Abelian models have comparatively quiet decay chains in the hidden
sector and the leptons will thus carry a higher fraction of the total
momentum of the event.

  An alternative possibility is that the connection between the
hidden sector and visible sector is via a singlet which has a
small coupling with the visible-sector Higgs fields and a large
coupling to hidden-sector Higgs fields.  For example, we can
consider the model of Section~\ref{singletmed} with the
superpotential coupling $W \supset \zeta S H_u H_d$ and $\zeta \ll
1$. This coupling is not necessary to communicate SUSY breaking,
but it is one possible way to ensure that the lightest axion has a
decay mode.  In this case, a small component of the neutralino is
the singlino, which will again lead to the decay $\chi^0
\rightarrow H \tilde{H}$ and similar signatures as before.

\subsubsection{Lower-Energy $e^+ e^-$ Colliders}

  Lower-energy $e^+ e^-$ machines, such as Belle, BaBar, DA$\Phi$NE, KLOE
and CLEO, offer high-luminosity precision tests of these low-mass
hidden sectors by production of the hidden-sector particles
through mixing of the $U(1)_x$ gauge boson with the photon.
Potentially interesting processes include $e^+ e^- \rightarrow
\gamma + hidden$ where the hidden states have invisible decay
modes and the photon arises from ISR, $e^+ e^- \rightarrow hidden$
with subsequent decays of the hidden particles producing SM final
states, as well as hidden decays of SM resonances, such as the
$\Upsilon(1S)$. We discuss signatures in both gauge- and
singlet-mediated models, leaving more detailed studies for future
work.  For related recent studies of hidden-sector $e^+e^-$
collider signatures, see
Refs.~\cite{McElrath:2005bp,Borodatchenkova:2005ct,Batell:2009yf,Essig:2009nc,Reece:2009un}.

  When there is gauge kinetic mixing of the $U(1)_x$ with electric charge,
hidden-sector states are produced by the $s$-channel exchange of
the $U(1)_x$ gauge boson, $e^+ e^- \rightarrow X \bar{X}$, or
through a $t$-channel electron exchange, $e^++e^-\to \gamma\,Z_x$.
The related production cross-sections are given in
Refs.~\cite{Batell:2009yf,Essig:2009nc}, which find that for
GeV-scale mediators and a kinetic mixing of $\epsilon \sim
10^{-3}$, production cross-sections at low-energy $e^+ e^-$
machines are typically in the fb range. Given the ab$^{-1}$
collected at the $B$ factories, this implies hundreds to thousands
of hidden-sector particles have been produced at these machines.
Search modes, and the corresponding constraints from existing
searches, then depend on the decay chains of the hidden
particles~\cite{Batell:2009yf,Essig:2009nc}. The simplest
signatures result when the hidden states remain stable on collider
timescales, so that the search is simply for a photon plus missing
energy.  This type of general search remains to be done. As in the
previous discussion on hadron-collider signatures, in the cascade
decays of hidden Higgs and Higgsinos, $Z_x$ gauge bosons may be
radiated which decay to pairs of leptons with low invariant mass.

  The Belle and CLEO collaborations have already
put a strong constraint on these sectors by searching for
$\Upsilon(3S,2S) \rightarrow \Upsilon(1S)+\gamma \rightarrow
 hidden + \gamma$~\cite{Tajima,Rubin},
where the $\Upsilon(1S)$ decays to hidden particles via photon mixing.
Belle finds the stronger constraint
at $B(\Upsilon(1S) \rightarrow invisible) < 2.5 \times 10^{-3}$, which
should be satisfied as long as $\epsilon \lesssim 10^{-2-3}$.  These
bounds apply for both direct decays to the LHP, and heavier hidden-sector
states which are either meta-stable on collider time scales or decay
to states which are meta-stable.

  For the singlet-mediated models, any production of hidden-sector particles
must go through mixing of visible and hidden Higgses.  Because the
coupling of the Higgs to the initial state is very weak, constraints
from $e^+e^-$ colliders are also typically very weak.  A possible
exception is through the exotic decay
$\Upsilon \rightarrow \gamma + hidden$, as discussed in
\cite{Fayet}.  The $\Upsilon$ can decay to an (off-shell) MSSM
pseudoscalar Higgs and photon, with the MSSM pseudoscalar Higgs
mixing with a hidden Higgs, $h^c$ or $h$. The hidden Higgs may
then decay to two LHPs, resulting in a completely invisible decay
of the hidden Higgs.  Since $\zeta \ll 1$, however, these constraints
also turn out to be quite weak.

\section{Conclusions\label{concl}}

  As long as there is an asymmetry in the way that various particles
feel supersymmetry breaking, it is generic to end up with sectors
at hierarchically different mass scales.  This seems particularly
likely when supersymmetry breaking is communicated to the visible
sector through gauge interactions, in which case particles that
are not charged under the messenger gauge group will receive
suppressed contributions to their soft masses.  If there are no
additional couplings between these particles and the visible
sector, one generically expects the mass scale of the hidden
sector to be set by the gravitino mass.  On the other hand, it is
easy for additional mediator fields (e.g., high-scale fields
charged under both visible and hidden gauge groups) to feed
supersymmetry breaking into the hidden sector through
loop-suppressed contributions, giving rise to GeV-scale hidden
sectors.

Perhaps one of the simplest ways to add a GeV-scale hidden sector
is to consider a new $U(1)_x$ gauge group and a vector-like pair
of fields charged under it.  If the $U(1)_x$ couples to the
visible sector through kinetic mixing with hypercharge,
SUSY breaking is communicated to the hidden sector
through the kinetic mixing. Even in the {\em absence} of kinetic
mixing, viable scenarios can arise when a singlet couples to both
the SUSY-breaking and hidden sectors, but with a
suppressed coupling to the hidden sector.

  Some care must be taken in the construction of these models,
however.   For example, if the gravitino is lighter than the
states in the hidden sector, there is always a danger of
quasi-stable states which either decay after BBN or have too large
of a relic density.  In this case, viable scenarios arise only for
a very low SUSY-breaking scale or if higher-dimensional
operators allow for additional decays.  This situation is
generally alleviated if the gravitino is heavier than the lightest
hidden-sector particle, provided it retains a large enough
annihilation channel so that its relic abundance isn't too large.
This light stable particle is typically a good candidate for the
dark matter, though its mass is much less than the weak scale.   A
particularly interesting scenario arises
when supergravity effects are sequestered and $m_{3/2} \sim
m_{hid}$ -- the lightest fermion is kinematically not allowed to
decay to a gravitino and photon, but can still annihilate to
hidden-sector states which then decay.   This scenario is viable
when the supersymmetry breaking is communicated to the hidden sector
either via kinetic mixing or via a singlet.

  New particles and forces at the GeV scale may also be relevant
for explaining some of the recent
possible hints for dark matter seen by the DAMA, PAMELA, ATIC, and
PPB-BETS experiments.  The annual modulation signal seen at DAMA
can potentially be explained by the elastic scattering of a
GeV-scale component of dark matter, or by heavier dark matter scattering 
inelastically to a state that is heavier by $\sim 100 \kev$, with the 
new sector at the GeV scale naturally inducing the splitting.
GeV-scale gauge bosons can
also give rise to a Sommerfeld enhancement of the dark matter
annihilation today, giving a possible explanation for the excess
in electrons and positrons reported by PAMELA, ATIC, and PPB-BETS.
In addition, these GeV-scale hidden sectors can potentially be
probed at both future and present hadron and $e^+e^-$ colliders.

Whether these sectors are responsible for the signals observed 
by the recent results from the dark matter experiments, the 
presence of such sectors can be quite generic, and the dynamics 
and signatures very different from MSSM phenomenology.  Dark 
matter experiments, as well $e^+e^-$ and hadron colliders, can 
then give us many possible windows into the hidden world.

\section*{Acknowledgements}

We thank Brian Batell, Doug Finkbeiner, Jonathan Heckman, Dan Hooper, 
Markus Luty, John Mason, Peter Ouyang, Frank Petriello, Matt Schwartz, 
Tracy Slatyer, and Tomer Volansky for helpful discussions.
This work is supported in part by the Harvard Center for the
Fundamental Laws of Nature, by NSF grant PHY-0556111 (DP),
and by the DOE including grant DE-FG02-95ER40896 (KMZ).


\appendix

\section{Appendix: Kinetic Mixing Mediation\label{appa}}

  Supersymmetry breaking in the visible (MSSM) sector can be
mediated to an Abelian $U(1)_x$ sector by gauge kinetic mixing.
In order to study these effects, we use the method of analytic
continuation into superspace~\cite{Giudice:1997ni,ArkaniHamed:1998kj}.
For this, we determine the dependence of the running hidden-sector
gauge couplings $g_a(\mu)$ and wavefunction factors $Z_i(\mu)$ on the gauge
messenger mass scale $M$, and then promote $M$ to a chiral superfield $X$
or a real superfield $\sqrt{X^{\dagger}X}$.  The result of this
procedure represents the leading term in an expansion in $|F/M^2|$
of the soft terms, with the full result obtainable from a
diagrammatic analysis~\cite{Martin:1996zb}.

  It is convenient to work in the holomorphic basis where the
gauge couplings appear in front of the gauge kinetic terms and are
promoted to chiral superfields.  In this basis, the effective
Lagrangian at scale $\mu$ takes the form
\begin{eqnarray}
\mathscr{L} &=& \int d^2 \theta \left[ \frac{1}{4 g_Y^2(\mu)}
B^{\alpha} B_{\alpha} + \frac{1}{4 g_x^2(\mu)} X^{\alpha}
X_{\alpha} + \frac{\epsilon_h(\mu)}{2} B^{\alpha} X_{\alpha} \right] +
h.c. \\
&& + \int d^4 \theta \left[ Z_{i}(\mu) \phi^{\dagger}_i
e^{x_{i} V_x} \phi_i + ... \right],
\end{eqnarray}
Note that $\epsilon_h$ is related to the kinetic
mixing in the canonically normalized basis at leading order as
\beq\epsilon(\mu) =
\epsilon_h(\mu) g_Y(\mu)g_x(\mu).
\eeq
The exact RG equations for the holomorphic
gauge couplings are
\begin{equation}
\frac{d}{d t} \left( \frac{1}{g_a^2} \right) =
\frac{b_a}{8\pi^2},~~~~~a=x,\,Y,
\end{equation}
where $b_a = -\sum_i q_i^a q_i^a$ is the beta function coefficient
(and $q_i^a$ denotes the charge of field $i$).
The holomorphic-basis kinetic mixing runs according to
\beq
\frac{d}{dt}\epsilon_h = \frac{b_{xY}}{8\pi^2}
\eeq
with $b_{xY} = -\sum_ix_iY_i$.  From this, we see that: \emph{if there are
no fields charged under both $U(1)_x$ and $U(1)_Y$, the holomorphic
kinetic mixing does not run at all}.
Upon transforming to the canonical basis with the kinetic mixing
eliminated, we reproduce the RG equations listed in Ref.~\cite{Babu:1996vt}.

  Let us assume the gauge messengers are charged only under
the SM gauge groups, and that there are no fields at the messenger
scale charged under both hypercharge and the $U(1)_x$ group.
In this case, only $b_Y$ (but not $b_x$ or $\epsilon_h$)
changes across the messenger threshold, only $g_Y(\mu)$
depends on the messenger mass, and only a $U(1)_Y$
gaugino mass is generated in this basis.  The gaugino mass matrix
in the basis with explicit kinetic mixing is then simply
\beq
M_{gaugino} = \left(
\begin{array}{cc}
M_1&0\\
0&0
\end{array}\right).
\eeq The physical gaugino masses can receive higher loop
corrections through the messenger mass dependence of the
wavefunction renormalization $Z_{i}$, since it contributes to the
physical gauge couplings through the rescaling anomaly associated
with going into a canonical basis for the kinetic terms.  This
correction to $M_{x}$ is tiny, however, and will be of order $\sim
\frac{\epsilon_h^2}{(16\pi^2)^3} \frac{F}{M}$.

  We turn next to the scalar masses in the light hidden sector.
The one-loop anomalous dimension of a hidden field $\phi_i$
with $U(1)_x$ charge $x_i$ is given by
\begin{eqnarray}
\label{gammaphii}
\frac{d \ln Z_{i}}{d t} &=& \frac{x_{i}^2}{4 \pi^2}
\frac{g_{x}^2}{1-g_x^2g_Y^2\epsilon_h^2} \\
&\simeq& \frac{x_{i}^2}{4 \pi^2} \left[ g_x^2(\mu) +
\epsilon_h^2\,g_Y^2(\mu,M) g_x^4(\mu) + O(\epsilon_h^4) \right].\nnmb
\end{eqnarray}
This can be obtained by resumming double insertions of $\epsilon_h$
on gauge boson propagators in the mixed kinetic basis, or by transforming
to a basis with canonical kinetic terms and no explicit kinetic mixing.
%
%
Only the second term in the last line of Eq.~\eqref{gammaphii}
depends on the messenger mass, so we see immediately that the
squared scalar masses are suppressed by $\epsilon_h^2$.
Since this contribution to the anomalous dimension is proportional
to that of the right-handed selectron (normalized to have hypercharge $Y=1$),
we obtain
\begin{equation}
\left. \frac{\partial^2 \ln Z_{i}}{\partial M^2}\right|_{\mu =
M} = \epsilon_h^2 x_{i}^2
 g_x^4(M) \left. \frac{\partial^2 \ln Z_{E^c}}{\partial M^2}\right|_{\mu =
M}.
\end{equation}
From this we can simply read off the soft masses generated
at the messenger scale:
\begin{equation}\label{softmasses}
m_{i}^2(M) = \epsilon_h^2 x_{i}^2 g_x^4(M) m_{E^c}^2(M) =
\epsilon^2(M)  \frac{x_{i}^2 g_x^2(M)}{g_Y^2(M)} m_{E^c}^2(M).
\end{equation}
We stress that these results are quite general, in that they do
not depend on the details of the messenger sector.
Furthermore, this result can also be derived by considering the
two-loop graph for the scalar $\phi_i$ communicating
to the messengers through the kinetic mixing term.

Finally, as in gauge mediation to the visible sector,
hidden-sector A-terms and B-terms can be generated at the two-loop
level in the presence of hidden-sector trilinear or bilinear
couplings. These terms will be generated at the messenger scale at
order $\sim \frac{\epsilon^2}{(16\pi^2)^2} \frac{F}{M}$, and will
be subdominant relative to the RG effects discussed in
Section~\ref{rgeeffects}.

\section{Appendix: Renormalization Group Equations}

  We collect here the one-loop renormalization group equations
for the soft terms in the models considered in the text.
Throughout, we implicitly shift the visible- and hidden-sector
scalar soft masses by explicit FI terms such that $Tr(Ym^2)=Tr(x
m^2)=0$~\cite{Jack:1999zs}.

\subsection{Minimal $\mu'$ Model}

  The only hidden-sector interactions within this model
are the $U(1)_x$ gauge interactions.  This leads to the
RG equations
\bea
(4\pi)^2\frac{dm_{H^{(c)}}^2}{dt} &=& -8\,\sep^2\,x_H^2\,g_x^2\,|M_1|^2\\
(4\pi)^2\frac{d(B\mu')}{dt} &=& -4\,x_H^2\,g_x^2\,(B\mu') -
8\,\sep^2\,x_H^2\,g_x^2\,(M_1\,\mu').
\eea

\subsection{Hidden Sector NMSSM}

  In addition to gauge interactions, there is now a Yukawa interaction
with coupling $\lambda$ (and potentially a singlet self-coupling $\kappa$).
This leads to the RG equations
\bea
(4\pi)^2\,\frac{d\ln\lambda}{dt} &=& 3\,|\lambda|^2 + 2\,|\kappa|^2
-4\,x_H^2\,g_x^2\\
(4\pi)^2\,\frac{d\ln\kappa}{dt} &=& 3\,|\lambda|^2+6\,|\kappa|^2\\
(4\pi)^2\,\frac{dm_{H^{(c)}}^2}{dt} &=&
2\,|\lambda|^2(m_{H}^2+m_{H^c}^2+m_S^2+|A_{\lambda}|^2)
-8\,\sep^2\,x_H^2\,g_x^2\,|M_1|^2\\
(4\pi)^2\,\frac{dm_S^2}{dt} &=&
2\,|\lambda|^2(m_{H}^2+m_{H^c}^2+m_S^2+|A_{\lambda}|^2)
+ 4\,|\kappa|^2(3\,m_s^2+|A_{\kappa}|^2)\\
(4\pi)^2\,\frac{dA_{\lambda}}{dt} &=&
6\,|\lambda|^2\,A_{\lambda}+ 4\,|\kappa|^2\,A_{\kappa}- 8\,\sep^2\,x_H^2\,g_x^2\,M_1\\
(4\pi)^2\,\frac{dA_{\kappa}}{dt} &=&
12\,|\kappa|^2\,A_{\kappa}+6\,|\lambda|^2\,A_{\lambda}.
\eea



\end{document}